\documentclass[12pt,preprint]{aastex}
\usepackage{epsfig}
\usepackage{graphicx}

\def\ltsima{$\; \buildrel < \over \sim \;$}
\def\lsim{\lower.5ex\hbox{\ltsima}}
\def\gtsima{$\; \buildrel > \over \sim \;$}
\def\gsim{\lower.5ex\hbox{\gtsima}}
\newcommand{\LCDM}{$\Lambda$CDM}
\newcommand{\HST}{{\it HST}}

\begin{document}
\shorttitle{The galaxy luminosity function}
\shortauthors{Dahlen et al.}
\title{The Evolution of the Optical and Near-Infrared Galaxy 
Luminosity Functions and Luminosity Densities to $z\sim2$}
\author{
Tomas Dahlen\altaffilmark{1,2}, 
Bahram Mobasher\altaffilmark{1,3},
Rachel S. Somerville\altaffilmark{1},
Leonidas A. Moustakas\altaffilmark{1},
Mark Dickinson\altaffilmark{4}, 
Henry C. Ferguson\altaffilmark{1}, 
and Mauro Giavalisco\altaffilmark{1} 
}
\altaffiltext{1}{Space Telescope Science Institute, Baltimore, 
MD 21218; dahlen@stsci.edu,
mobasher@stsci.edu, somerville@stsci.edu, leonidas@stsci.edu, 
ferguson@stsci.edu, mauro@stsci.edu}
\altaffiltext{2}{Department of Physics, Stockholm University, 
AlbaNova University Center, SE-106 91 Stockholm, Sweden}
\altaffiltext{3}{Affiliated with the Space Telescope Division of 
the European Space Agency,
ESTEC, Noordwijk, Netherlands.}
\altaffiltext{4}{National Optical Astronomy Observatory, P.O. 
Box 26732, Tucson, AZ 85726; med@noao.edu}

\begin{abstract}
Using Hubble Space Telescope and ground-based $U$ through $K_s$
photometry from the Great Observatories Origins Deep Survey, we
measure the evolution of the luminosity function and luminosity density
in the rest-frame optical ($UBR$) to $z\sim2$, bridging the poorly 
explored ``redshift desert'' 
between $z\sim1$ and $z\sim2$. We also use deep near-infrared observations
to measure the evolution in the rest-frame $J$-band to $z\sim1$.
Compared to local measurements from
the SDSS, we find a brightening of the characteristic magnitude, $M^*$, 
by $\sim2.1$, $\sim0.8$~and $\sim0.7$~mag between $z\sim0.1$~and $z\sim1.9$,
in $U$, $B$, and $R$, respectively.
The evolution of $M^*$~in the $J$-band is in the opposite sense, 
showing a dimming between redshifts $z\sim 0.4$~and $z\sim 0.9$ . 
This is consistent with a scenario in which the mean star formation 
rate in galaxies was higher in the past, while the mean stellar mass 
was lower, in qualitative agreement with hierarchical galaxy formation models.
We find that the shape of the luminosity function is strongly
dependent on spectral type and that there is strong evolution with
redshift in the relative contribution from the different spectral
types to the luminosity density.  
We find good agreement with previous measurements, supporting an increase in 
the $B$-band luminosity density by a factor $\sim2$~between the local value 
and $z\sim1$, and little evolution between $z\sim1$~and $z\sim2$. We
provide estimates of the uncertainty in our luminosity density
measurements due to cosmic variance.  We find good agreement in
the luminosity function derived from an $R$-selected and a
$K_s$-selected sample at $z\sim1$, suggesting that optically selected
surveys of similar depth ($R\la24$) are not missing a significant
fraction of objects at this redshift relative to a near-infrared-selected
sample. We compare the rest-frame $B$-band luminosity functions from 
$z$~0--2 with the predictions of a semi-analytic hierarchical model of 
galaxy formation, and find qualitatively good agreement. In particular, 
the model predicts at least as many optically luminous galaxies 
at $z\sim$~1--2 as are implied by our observations.
\end{abstract}


\keywords{Galaxies: distance and redshifts -- Galaxies: evolution -- Galaxies:
fundamental parameters -- Galaxies: high-redshift -- Galaxies: luminosity function}

\section{Introduction}
\label{sec:intro}
The luminosity function (hereafter LF) is one of the most fundamental
metrics of the galaxy population, and is essential for characterizing
statistical properties of galaxies and their evolution. Studying the
LF as a function of cosmic epoch, galaxy type, and environment
provides insights into the physical processes that shape galaxies. The
measured parameters of the LF --- the normalization, faint-end slope,
and location of the characteristic luminosity or ``knee'' --- are
strong constraints on galaxy formation models and indeed are often
considered the most basic test of a models' viability. When integrated
over all luminosities, the LF provides a measure of the global
luminosity density. The dependence of this quantity on wavelength and
cosmic time allows us to probe the cosmic star formation and stellar
mass assembly rate.

To construct the LF, we require a complete and unbiased sample of
galaxies to a given flux limit, with well-known selection functions
and available redshifts. At any given redshift, the survey should
probe a large enough volume to control cosmic variance, and contain
enough galaxies in each magnitude bin to minimize Poisson noise.
Recently, these problems have been largely overcome for measures of
the local LF ($z\sim0$) by several wide-area, multi-waveband surveys
with follow-up spectroscopy.  Recent local surveys include the 2dF
Galaxy Redshift Survey (2dF-GRS; Colless et al. 2001; Norberg et
al. 2002), the Sloan Digital Sky Survey (SDSS; Stoughton et al. 2002;
Blanton et al. 2001, 2003) and the Two Micron All Sky Survey (2MASS;
Jarrett et al. 2000; Kochanek et al. 2001; Cole et al. 2001). These
studies have resolved the debate about the normalization and faint-end
slope of the local optical LF that persisted for many years (Marzke 
et al. 1994; da Costa et al. 1994; Lin et al. 1996).

In order to study the LF at high redshift, as is our goal here, the
desire for wide area coverage and large samples must be balanced
against the need for deep enough photometry to probe faint-wards of the
``knee'' in the LF ($L_*$). In addition, in order to construct the
high redshift LF in a given fixed \emph{rest wavelength}, we require
photometry at wavelengths long-wards of the desired rest
frame. Obtaining measures of the rest-frame optical LF at redshifts
above $z\sim1$ has therefore proved challenging, because of the
difficulty of obtaining deep near-infrared (hereafter NIR) photometry 
over large enough
fields. Obtaining spectroscopic redshifts for LF studies at high
redshift is also extremely challenging. Most previous studies of the
LF out to $z\sim1$ based on spectroscopic surveys, such as the Canada
France Redshift Survey (Lilly et al. 1995), CNOC2 (Lin et al. 1999)
and the Caltech Faint Galaxy Redshift Survey (Cohen 2002) represent
heroic efforts but were seriously limited by small number statistics
and uncertainties due to cosmic variance.  Two major efforts exploiting
multi-object spectrographs on large telescopes are directed at
obtaining redshifts for significant volumes out to $z\sim1.5$ and will
greatly improve this situation: the DEEP2 redshift survey on the Keck
telescope (Davis et al. 2003) and the VIMOS-VLT Deep Survey (VVDS; Le
F\`{e}vre et al. 2004).  See also the Team Keck Treasury Redshift Survey 
(Wirth et al. (2004).  However, even these surveys are optically
($R$-band) selected and may suffer incompleteness for faint and/or red
objects at $z\ga1$.

An alternative and powerful method of obtaining redshifts for
statistically large samples to deep flux levels is provided by the
photometric redshift method, analogous to low resolution
spectroscopy. Due to improved accuracy in photometric redshift
techniques in recent years 
($\Delta_z\equiv\langle|z_{\rm phot}-z_{\rm spec}|/(1+z_{\rm spec})\rangle\lsim 0.1$; 
Benitez et al. 2000; Mobasher et al. 2004), sufficiently reliable redshifts 
are now available for statistical studies of galaxies.  Large, deep
multi-waveband surveys using photometric redshifts to measure LFs to 
$z<1.5$~have recently been completed by, e.g., the Las Campanas Infrared 
Survey (Chen et al. 2003) and the COMBO-17 project (Wolf et al. 2003). 
To higher redshifts ($z<3-5$), but for relatively smaller areas, LFs have 
been presented by e.g.,the Subaru Deep Survey (Kashikawa et al. 2003), the 
FORS Deep Field (Gabasch et al. 2004) and by Poli et al. (2003).

In this paper, we make use of observations obtained as part of the
Great Observatories Origins Deep Survey (GOODS; Giavalisco et
al. 2004) in the Chandra Deep Field South (CDF-S), also known as the
GOODS-S field. We combine the results from a wider area ($\sim 1100$
arcmin$^2$), optically selected ($R_{AB}<24.5$) catalog with a smaller
area ($\sim 130$ arcmin$^2$) but very deep NIR selected ($K_{s,
AB}<23.2$) catalog to measure galaxy luminosity functions and the
luminosity density from $z\sim0.1$--2.  Using the optically selected
sample, we construct these quantities from $z\sim 0.1$--1 for the
overall galaxy population and for different spectral types. The NIR-selected 
catalog represents an unprecedented combination of area and
depth, allowing us to probe deeper down the rest-frame optical
luminosity function in the difficult ``spectroscopic desert'' regime
of $z\sim1$--2 than has been possible previously. We construct the
rest-frame optical LF to $z\sim2$~using the $K_s$-selected
catalog. We also use the $K_s$-selected catalog to determine the
evolution of the rest-frame $J$-band LF to $z\sim1$.
  
In the next Section, we present the multi-waveband dataset used in
this study and the technique used to estimate photometric
redshifts. The methods used to measure the LFs, incorporating
photometric redshift errors using the redshift probability
distributions, and our approach for estimating the uncertainty due to
cosmic variance are discussed in \S \ref{sec:methods}.  In \S
\ref{sec:results} we present our measurements of the LF from
$z\sim0.1$--2 for the global population and the LF divided by spectral
type, and compare our results with the predictions of a semi-analytic
model of galaxy formation. We use these results to derive the
integrated luminosity density from $z\sim0.1$--2. We discuss our
results and conclude in \S \ref{sec:discussion}.

Throughout this paper we adopt a flat cosmological constant dominated
cosmology ($\Omega_{\Lambda}=0.7, \Omega_M=0.3$) and a Hubble constant
$H_0$=70 km s$^{-1}$~Mpc$^{-1}$. All magnitudes are in the AB system
(Oke \& Gunn 1983).

\section{Multi-waveband data and photometric redshifts}
\label{sec:data}

The photometric data for the CDF-S were obtained using ground-based
optical and NIR observations from ESO (2.2m WFI, VLT-FORS1,
NTT-SOFI and VLT-ISAAC) and CTIO (4m telescope) covering $UBVRIJHK_s$,
as well as space-based Hubble Space Telescope (\HST) Advanced Camera
for Surveys (ACS) observations in $BViz$.  A summary of the
observations and data reduction is given in Giavalisco et
al. (2004). For the purpose of this study, two different photometric
catalogs were produced. Using SExtractor (Bertin \& Arnouts 1996), we
constructed a WFI $R$-band selected catalog including all ground-based
photometry, and an ISAAC $K_s$-band selected catalog including ISAAC
$JHK_s$~and ACS $BViz$ photometry. The $R$-selected catalog covers
$\sim 1100$~square arcmin while the $K_s$-selected catalog covers
$\sim 130$~square arcmin. Magnitudes used for calculating LFs are
based on SExtractor MAG-AUTO, while colors used for determining
photometric redshifts and spectral types are based on 3 arcsec
diameter aperture magnitudes. Images were convolved to a common
point-spread function (PSF) before measuring aperture magnitudes. The
full-width at half maximum (FWHM) for the WFI $R$-selected catalog is
$\sim0.9$~arcsec, while the ISAAC $K_s$-selected catalog has a FWHM of
$\sim0.45$~arcsec.

The apparent magnitude limits are $R=24.5$~and $K_s=23.2$, resulting
in 18,381 galaxies at $0.1<z<1.0$~and 2,768 galaxies at $0.1<z<2.0$~for
the $R$- and $K_s$-selected catalogs, respectively.  The magnitude
limits are sufficiently bright ($\sim$~0.5 mag brighter than the
magnitude where the number counts start to deviate from a pure power
law) to allow accurate photometric redshifts estimates and yet faint
enough to make it possible to study the evolution of the faint-end of
the LF, as well as the integrated luminosity density, for a major part
of the galaxy population.

Using the available 13 ($R$-selected catalog) and 7 ($K_s$-selected
catalog) passbands, we estimate photometric redshifts for all galaxies
in this survey to the magnitude limits given above.  The photometric
redshift code developed for this investigation is based on
template-fitting method (e.g., Gwyn 1995; Mobasher et al. 1996) and
includes a Bayesian prior based on an input LF similar to the approach
used in, e.g., Kodama et al. (1999) and Dahlen et al. (2004).  At each
redshift, the absolute $V$-band magnitude of the object is calculated
and compared to an input LF.  The main effect of the absolute
magnitude prior is to discriminate between cases where the chi-square
fitting results in two probability peaks due to confusion between the
Lyman- and the 4000\AA-break.  The absolute magnitude that the object
would have at the two redshift peaks can discriminate between the
choices, i.e., absolute magnitudes being significantly brighter than
$M^*$~is regarded as increasingly improbable. This prior is similar to
the method of using a single bright magnitude cutoff for the absolute
magnitude to select against wrongly assigned photometric redshifts, an
approach used by e.g., COMBO-17 (Wolf et al. 2003).

In most cases, however, the prior does not affect the results and the
photometric redshift is given by the best chi-square fit.  Note also
that we use a flat faint-end slope for the input prior LF. This is
equivalent to not imposing any luminosity prior at all at faint
    magnitudes. This is important since we do not want to bias the slope
of the LF that we later measure using the results from the photometric
redshifts.  Absorption due to intergalactic HI is included using the
parameterization in Madau (1995).  We use the four different template SEDs,
consisting of E, Sbc, Scd, and Im, from Coleman et al. (1980), extended to UV
and IR wavelengths by Bolzonella et al. (2000). We also include two starburst
templates from Kinney et al. (1996, SB2 and SB3). Ten additional templates are
constructed by interpolating between subsequent SEDs.

Comparison between photometric and spectroscopic redshifts for
$\sim400$~galaxies in this sample results in an 
accuracy of $\Delta_z\sim 0.12$. 
After excluding a small number of outliers ($\sim$ 3\% with $\Delta_z>0.2$), 
the accuracy becomes $\Delta_z\sim 0.06$.

As part of the procedure for estimating the photometric redshift of
individual galaxies by fitting the observed SEDs to templates, we also
obtain the spectral type and the redshift probability distribution for
each galaxy. These SED types are used in the following sections to
study the evolution of the LFs for galaxies of different spectral
types, while the redshift probability distribution for each galaxy is
used to incorporate the errors in the photometric redshifts, which
propagate through the estimated LFs (\S \ref{sec:methods:lf}).

We divide the galaxies into three broad spectral types; early-types, 
late-types, and starbursts. Types with a best fitting SED dominated by the 
elliptical spectrum are defined as early-types, while galaxies dominated by 
one of the spiral galaxy spectra are called late-types. Starbursts are best 
fitted by one of the two starburst spectra. The approximate division between 
types in restframe color is that objects with $B-V>0.7$~are early-types and 
objects with $B-V<0.25$~are starbursts, while objects with intermediate color
are late-types.

Note that the {\em spectral types}, which we here divide into
early-types, late-types, and starbursts, represent the average color of the 
galaxies. These do not necessarily have a one-to-one
correlation to {\em morphological types}, e.g., morphological
ellipticals, spirals, and irregulars. At least at moderate redshifts,
however, the correlation appears to be strong.

\begin{figure*}
\centerline{
\psfig{figure=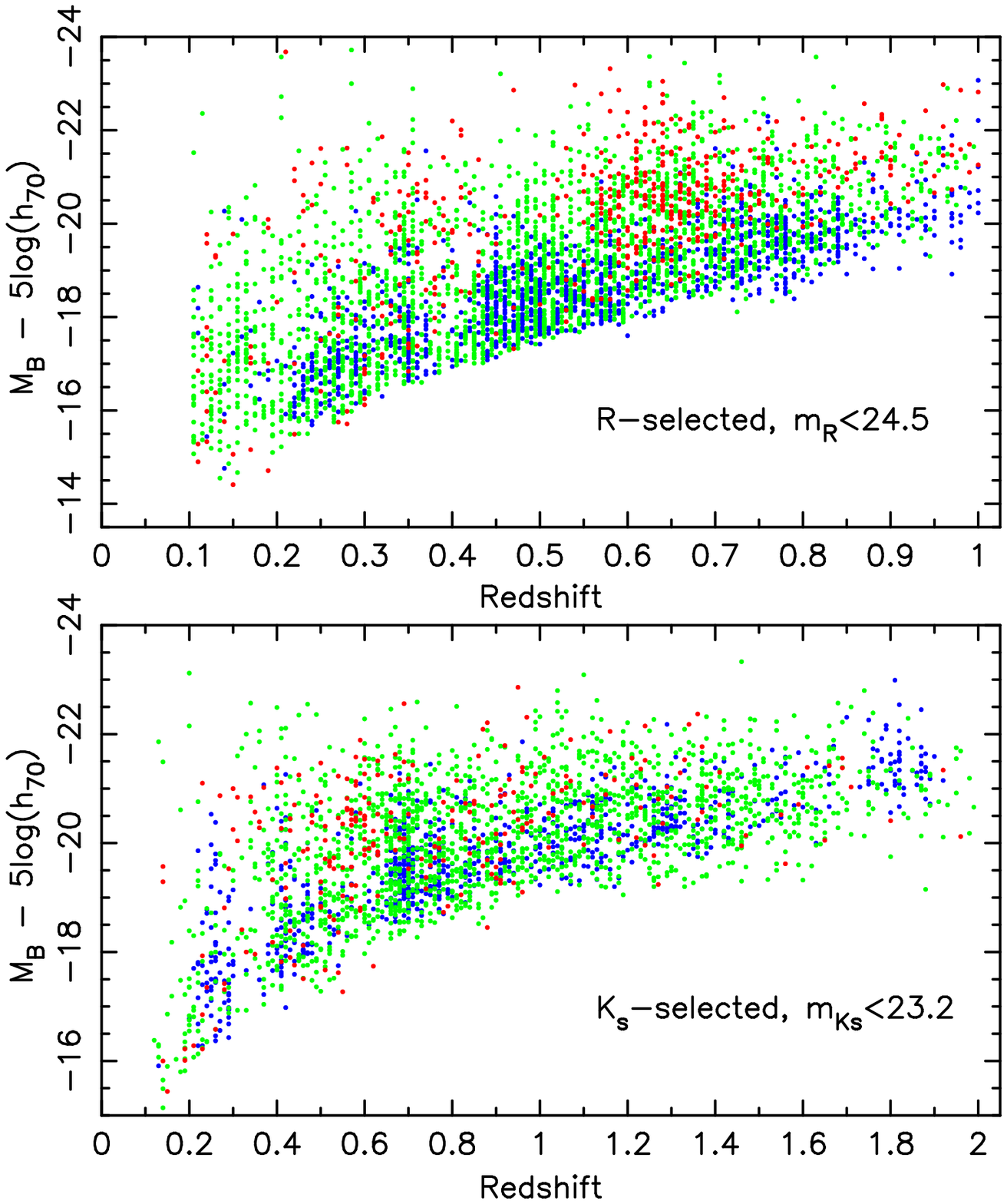,height=13.0cm}}
\caption{Rest-frame $B$-band magnitude vs. redshift for the $R$-selected 
sample (top panel) and $K_s$-selected sample (bottom panel). Galaxies with 
early-type, late-type and starburst spectral types are shown with red, green
and blue dots, respectively.  
\label{fig1}}
\end{figure*}

In Figure \ref{fig1}, we show the absolute magnitude vs. redshift 
relation for the $R$-band and $K_s$-band selected samples. The early-type, 
late-type and starburst galaxies are represented by red, green and blue color,
respectively. Absolute magnitudes are calculated according to the recipe 
described in the next Section.
\section{Method}
\label{sec:methods}

In this section we present the technique used to estimate the LF and
our procedure for accounting for the errors in our photometric
redshifts. We also describe our approach for estimating the
uncertainty in our results due to cosmic variance.

\subsection{Deriving the Luminosity Function}
\label{sec:methods:lf}

Using the information on photometric redshift and best-fitting SED
template, we calculate the rest-frame absolute magnitude in filter
$Y$, $M_Y$, from the observed apparent magnitude in filter $X$, $m_X$,
using the general equation
\begin{equation}
M_Y = m_X - 5 \log(D_L(z)/10\ {\rm pc}) - K_{XY} (z,T)
\end{equation}
where $K_{XY}(z,T)$~is the K-correction at redshift $z$~for template
type $T$, correcting from observed filter $X$, to rest-frame filter
$Y$, and $D_L(z)$ is the luminosity distance. A detailed description
of how we calculate rest-frame magnitudes and the luminosity distance
is given in the Appendix.

We use the $1/V_{\rm max}$ method (Schmidt 1968) to calculate the LF 
according to
\begin{equation}
\Phi (M)dM=\sum_i \frac{1}{V_i(M_i)}
\end{equation}
where ${V_i(M_i)}$ is the observable comoving volume in which a galaxy
$i$~with absolute magnitude $M_i$~is detectable, considering the
apparent magnitude and redshift limits of the survey.  The sum is
taken over all galaxies in the magnitude range $M-\Delta(M)/2<M_i<M+
\Delta(M)/2$. When evaluating equation (2), we set $\Delta M=0.5$
magnitudes.  The comoving volume for any given galaxy is given by the redshift
range $z_{i,{\rm min}}-z_{i,{\rm max}}$, where $z_{i,{\rm min}}={\rm
  Max}(z_{\rm low},z_{m+})$ and $z_{i,{\rm max}}={\rm
  Min}(z_{\rm high},z_{m-})$. Here $z_{\rm low}$~and $z_{\rm high}$~are the 
lower and upper redshift limits of the redshift bin where the LF is
determined and $z_{m+}$~and $z_{m-}$~ are the redshifts limits where a
galaxy with absolute magnitude $M_i$~will have an apparent magnitude
within the magnitude limits of the survey.  For a survey covering an
area $\Delta \Omega$, the volume becomes,
\begin{equation}
V_i=\frac{c\Delta
  \Omega}{H_0}\int_{z_{i,{\rm min}}}^{z_{i,{\rm max}}}\frac{D_L(z)^2}{(1+z)^2}
\end{equation}
 
$$\times (\Omega_M(1+z)^3 + \Omega_k(1+z)^2 + \Omega_{\Lambda})^{-1/2}dz\rm{,}$$

where $\Omega_k=1-\Omega_M-\Omega_{\Lambda}$, $H_0$~is the Hubble
constant, $c$~ is the speed of light and $D_L$~is the luminosity
distance (see Appendix).  The $1/V_{\rm max}$ method is does not assume 
any underlying parametric form for the LF. Instead, after determining 
the number of objects in each magnitude
bin (and their errors), it is possible to fit any desired functional
form to the data.  A drawback with the $1/V_{\rm max}$ method is
that systematic biases may be introduced by large scale fluctuations
and clustering within the observed field (de Lapparent et al. 1989).

Deriving the LF requires determination of the absolute magnitude of
each galaxy.  With redshifts estimated from photometric redshift
methods, the relatively large errors in redshifts propagate to errors
in the luminosity distances and K-corrections, and therefore also to 
the absolute magnitudes.

As faint
galaxies are much more abundant than bright ones, a larger fraction of
intrinsically faint galaxies is shifted to the bright end, than
intrinsically bright galaxies are moved to the faint-end (Chen et
al. 2003).  Thus, uncertainties in the redshifts may introduce biases
in the determination of the LF, and lead to systematic errors when
deriving the LF parameters, e.g., in determining the characteristic
luminosity and the faint-end slope.

To incorporate the redshift uncertainties, we use for each object its
corresponding redshift probability distribution, $p_i(z)$, instead of
a single redshift value. The probability distribution is derived from
the chi-square of the template fitting including input priors.

When summing the magnitude bins in equation (2), we use for each
object a redshift range $z_-<z<z_+$ where the lower limit, $z_-$,
corresponds to $M_i+\Delta M$ and the upper limit, $z_+$, corresponds
to $M_i-\Delta M$, as calculated using equation (1). The contribution
from each galaxy is weighted by the probability given by
\begin{equation}
P_i=\int^{z_+}_{z_-}p_i(z)dz.
\end{equation} 
This is expressed as
\begin{equation}
\Phi (M)dM=\sum_i \frac{P_i}{V_i(M_i)}
\end{equation}

Alternatively, the LF can be derived using a maximum likelihood (ML) approach 
(Sandage et al. 1979).  An advantage with the ML method is that it is not
as sensitive to large scale fluctuations as the $1/V_{\rm max}$ method.  On 
the other hand, the ML method requires fixing the functional form of the LF 
which can be misleading since this form is not in general known a priori. 
Analogous to the method described above, Chen et al. (2003) describes a
method for incorporating photometric redshift errors when deriving the LF
using the ML method.

\subsection{Cosmic variance}
\label{sec:methods:cosvar}

An important source of uncertainty in estimates of galaxy densities
and related quantities (LF, luminosity density, etc) in deep fields
is the field-to-field variation due to large scale structure ---
commonly referred to as \emph{cosmic variance}. It is generally
difficult to estimate the magnitude of this variance empirically, as
this requires measures of galaxy clustering on scales larger than the
field in question and these are generally not available. In Somerville
et al. (2004a), we presented an approach for estimating the cosmic
variance for observed populations based on the expectations of
clustering in the Cold Dark Matter (CDM) theory and a simple model for
galaxy bias. These results were appropriate for the variance in the
\emph{number counts} (or number densities) of galaxies discussed in
that paper, but are not directly applicable to the variance in the
\emph{luminosity density}, which is of interest here. Because more
luminous galaxies presumably occupy more massive dark matter halos,
and halo clustering is known to be a strong function of mass in the
CDM theory, we must account for this in estimating the cosmic variance
for mass- or luminosity-weighted quantities such as the luminosity
density. We have developed a simple model to address this problem,
which we briefly outline here. More details and comparisons with
numerical simulations will be given in a future work (Newman \&
Somerville, in preparation).

First, we compute the fractional variance for the underlying dark
matter density field in each redshift bin used in our analysis. We
approximate the bin geometry as a rectangular solid, as in Newman \&
Davis (2002), and use the power spectrum for a standard ($n=1$) \LCDM\
cosmology with the same parameters assumed throughout this paper and a
normalization $\sigma_8=0.9$. The resulting values of $\sigma_{\rm
DM}$ are given in Table~\ref{tab:cosvar} for both the WFI
($R$-selected) and ISAAC ($K_s$-selected) catalog geometries.

Next, we estimate an effective rest $B$-band \emph{luminosity weighted
effective bias} in the following manner. We assume that the ratio of
total halo mass to total $B$-band light as a function of halo mass is
given by the functional form
\begin{equation}
\left< \frac{M}{L} \right>(M) = 0.5 \left( \frac{M}{L} \right)_0
\left[ \left(\frac{M}{M_c}\right)^{-\gamma_1} +
  \left(\frac{M}{M_c}\right)^{-\gamma_2} \right]
\end{equation}
following van den Bosch et al. (2003), and adopt the values for
the parameters $\left( \frac{M}{L} \right)_0$, $M_c$, $\gamma_1$ and
$\gamma_2$ from their Model A. van den Bosch et al. (2003) showed that
this model reproduces the luminosity function and luminosity dependent
clustering of $B$-band selected galaxies observed locally by the
2dF-GRS. We then define the luminosity-weighted effective bias as
\begin{equation}
b_{\rm lum} = \frac{1}{\rho_L} \int_{M_{\rm min}} \frac{dn}{dM}(M,z)
\left< \frac{M}{L} \right>(M)\, M dM
\end{equation}
where $\rho_L$ is the total luminosity density and
$\frac{dn}{dM}(M,z)$ is the dark matter halo mass function at redshift
$z$. We adopt $M_{\rm min}=10^{10} M_{\odot}$. As the light-to-mass
ratio declines rapidly with decreasing halo mass, the results do not
depend sensitively on $M_{\rm min}$ as long as it is of this order or
smaller. The fractional root variance in the luminosity density is
then simply $\sigma_L = b_{\rm lum} \sigma_{\rm DM}$.

Note that we assume that the relationship between light and mass
$\left< \frac{M}{L}\right>(M)$ does not change with redshift. This is
equivalent to assuming that the \emph{halo occupation function} does
not change with time, so that all redshift dependence in our model is
contained in the changing dark matter halo mass function and
clustering amplitude rather than the way that galaxies of a given
luminosity trace the underlying dark matter halos (e.g., Moustakas \&
Somerville 2002). It has been shown
that this assumption is consistent with observations out to redshifts
$z\sim1$ (Yan et al. 2003; Coil et al. 2004), and the
relatively constant luminosity density from $z\sim1$--2 that our
observations imply is also consistent with this assumption. Luminous
($L\gsim L_*$) galaxies must be positively biased ($b>1$) in the \LCDM\
model considered here, so the variance for the dark matter
$\sigma_{\rm DM}$, probably underestimates the luminosity-weighted
variance. If there is evolution in $\left< \frac{M}{L}\right>(M)$, it
is likely to be in the sense that less massive halos will contribute a
higher fraction of the total luminosity at high redshift than locally
(as is found in semi-analytic models of galaxy formation),
which would tend to \emph{decrease} our estimate of $b_{\rm
lum}$. Therefore, the true variance is almost certainly bracketed by
$\sigma_{\rm DM}$ and $\sigma_{L}$. We provide both values in
Table~\ref{tab:cosvar}. We find that the 1-$\sigma$ uncertainty in the
dark matter density due to cosmic variance for the WFI ($\sim1100$
arcmin$^2$) fields is $\sim 15$ percent, while the uncertainty on the
luminosity density is $\sim16$--19 percent. For the smaller ISAAC
field ($\sim130$ arcmin$^2$), the DM uncertainty is $\sim 15$--30
percent, while the luminosity weighted values are $\sim 16$--45
percent.

We also note that the empirical relationship between luminosity and DM
halo mass that we have adopted here is specific to the $B$-band. We
expect that this relationship will be qualitatively similar, but
perhaps different in detail, in other wavebands. The redshift
evolution may also be more pronounced in shorter wavebands (e.g. the
$U$-band). Therefore we do not present cosmic variance estimates for
the other wavebands, but expect that the estimates given for the
$B$-band should be fairly representative for the other bands.

\section{Results}
\label{sec:results}

\subsection{Luminosity Functions}
\label{sec:results:lf}

In this investigation, we use the $1/V_{\rm max}$~technique to calculate
the LF, which is thereafter fitted to a Schechter function (Schechter 
1976).  We calculate the rest-frame LFs in the $U$- and $B$-bands using 
both the $R$-selected (WFI based) and $K_s$-selected (ISAAC based) catalogs, 
and the rest $R$- and $J$-band using only the $K_s$-selected catalog. We adopt
two sets of redshift bins appropriate to the volume and depth of each
catalog. The WFI-based rest-frame $U-$, $B-$ and ISAAC-based $J$-band
LFs are determined in three redshift intervals, $0.1<z<0.5$,
$0.5<z<0.75$ and $0.75<z<1.0$. The redshifts corresponding to the
volume mid-point of each bin are {\it \~{z}}=0.39, {\it \~{z}}=0.64 and 
{\it \~{z}}=0.88, respectively (hereafter we use '{\it \~{z}}' to denote this
redshift). For the optical bands, we further derive the 
type-dependent LFs in each bin. For the $J$-band, we do this only for the
lowest redshift bin where statistics are sufficient.  Using the
$K_s$-selected catalog, we extend the determination of the rest-frame
$U$-, $B$-~and $R$-band LF to $z\sim2$, using 6 bins with approximately equal
comoving volume between $z=0.1$ and $z=2$. The redshift of the volume 
mid-points of these bins are {\it \~{z}}=0.62, {\it \~{z}}=0.98, 
{\it \~{z}}=1.24, {\it \~{z}}=1.48, {\it \~{z}}=1.69 and {\it \~{z}}=1.90, 
respectively. 

\subsubsection{`Quasi-local' LF}
\begin{figure*}
\centerline{
\psfig{figure=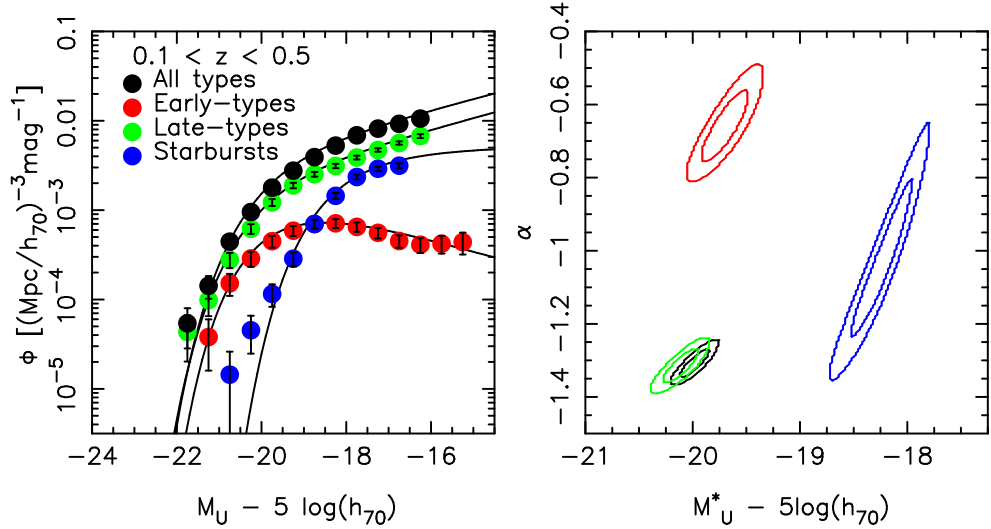,height=7.0cm}}
\caption{The left panel shows the rest-frame `quasi-local' $U$-band
  luminosity function in the redshift range 0.1$<z<$0.5. Black dots
  show the composite LF, while red, green, and blue dots show the
  type-specific LFs for early-types, late-types, and starbursts,
  respectively. The types are determined from best fitting spectral
  templates. The right panel shows 68.3 and 95 \% error ellipses for
  $M^*$ and $\alpha$~from Schechter function fits to different
  populations.
\label{fig2}}
\end{figure*}

\begin{figure*}
\centerline{
\psfig{figure=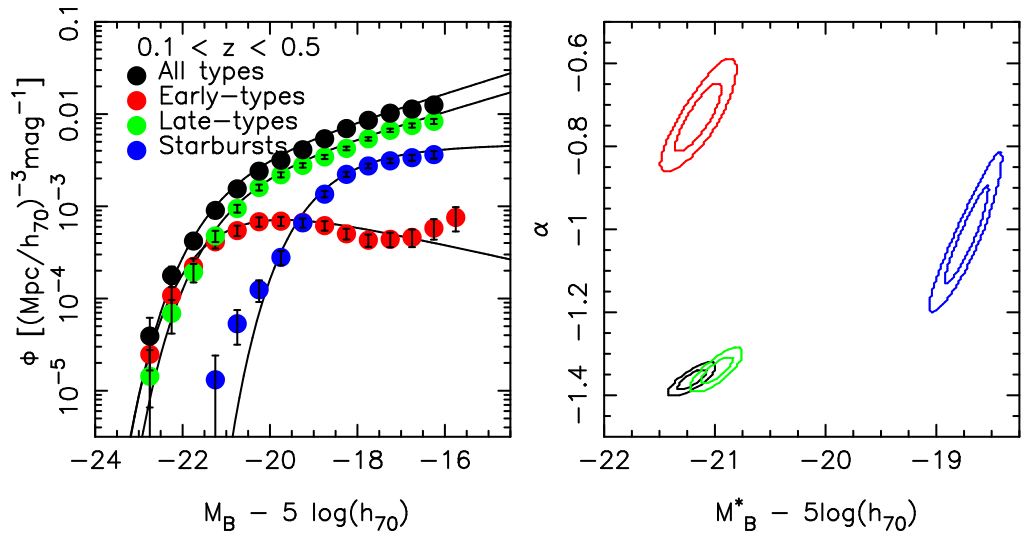,height=7.0cm}}
\caption{The rest-frame `quasi-local' $B$-band luminosity
  function. Symbols are the same as in Figure \ref{fig2}.
\label{fig3}}
\end{figure*}

\begin{figure*}
\centerline{
\psfig{figure=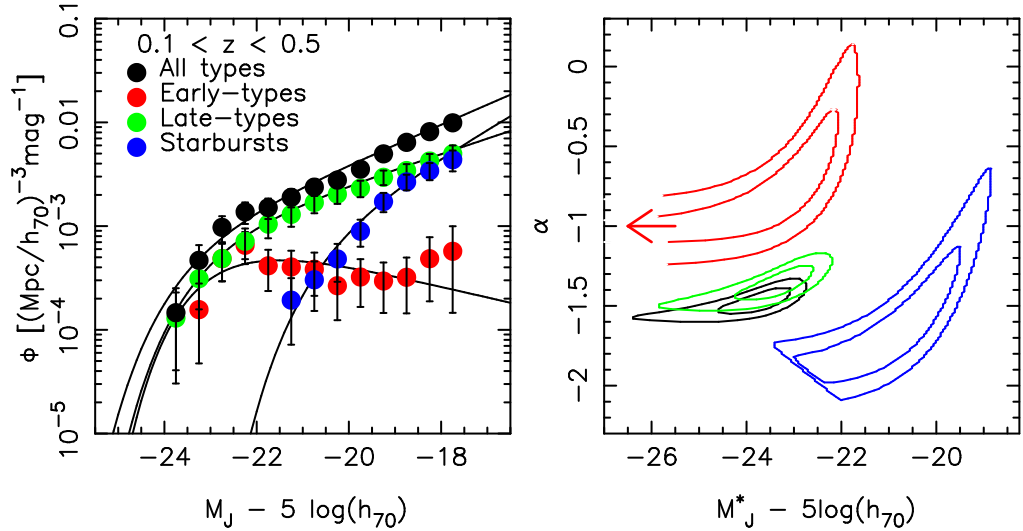,height=7.0cm}}
\caption{The rest-frame `quasi-local' $J$-band luminosity
  function. Symbols are the same as in Figure \ref{fig2}. Note that
  $M^*$ for the early-type population is not constrained at bright
  magnitudes since a Schechter function does not well represent the LF
  (see text for details).
\label{fig4}}
\end{figure*}

In the left panels of Figures \ref{fig2}-\ref{fig4}, we show the composite 
rest-frame $U-$, $B-$ and $J$-band LFs, as well as the type-specific LFs 
divided into three spectral types: early-types, late-types, and starbursts.
The best-fitting Schechter LF parameters ($M^*$, $\alpha$, and
$\phi^*$) are listed in Tables $2-4$. The error ellipses for the
Schechter function parameters for both the total LFs and the different
spectral type LFs are presented alongside the LFs.  These contours
represent 68.3\% and 95\% confidence intervals and correspond to
$\Delta \chi^2=2.3$~ and $\Delta \chi^2=6.0$~above the best-fitting
$\chi^2$, respectively.

The shape of the LFs varies dramatically between different spectral
types.  While the late-type LF mainly follows the total LF, showing a
fairly steep faint-end slope, the early-type LF has a Gaussian shape
which is significantly flatter at faint magnitudes. There is also 
an indication of an upturn of the LF at the very faintest magnitudes.
We discuss this below. The starburst-type
population also has a fairly steep faint-end slope, with number
densities of these galaxies at the bright-end much smaller (by an
order of magnitude) than other types.

The composite and type-dependent LFs in the $U$- and $B$-bands are
similar, both in terms of the shape and faint-end slopes.  There is a
difference in that the early-type population is more dominant in
$B$~than in $U$, while the opposite is true for the starburst-type
population. This difference is more evident in the next Section where
we derive the fractional contribution to the luminosity density from
different spectral types.

The composite rest-frame $J$-band LF in Figure \ref{fig4}, has an
overall shape similar to its optical counterparts. Also, the type-specific 
LFs have shapes similar to the optical.  Compared to the
optical, especially the $U$-band, the starburst population is
relatively fainter in $J$, with the characteristic magnitude of the
starburst population being $\sim3.0$~mag fainter than the composite LF
in $J$.  In the $U$- and $B$-band, the difference between the
characteristic magnitude of the starburst-type LF and the composite LF
is $\sim1.8$~mag and $\sim2.5$, respectively.  The early-type
population is also more dominant in $J-$~than in $U$-band. This
difference is not significant when comparing the $B$- and
$J$-bands. These trends are also reflected in the contribution by
different populations to the total luminosity density in different
bands, presented in the next section.

At the faint-end of the early-type LF, we note a possible up-turn of
the LF in all bands (most evident in $B$). A similar result was seen in the 
LF obtained for early-spectral-type galaxies from the 2dF-GRS
(Madgwick et al. 2002).  This up-turn may be an indication
of an abundant population of faint early-type galaxies that starts to
dominate the early-type LF at faint magnitudes.  This may be analogous
to the dwarf elliptical population that starts to dominate over normal
ellipticals in local cluster of galaxies and galaxy groups (e.g.,
Binggeli et al. 1988).  An alternative explanation could be
contamination from faint red M-stars, whose SEDs are similar to
early-types and may cause confusion, especially at faint magnitudes
where it is more difficult to classify stars using e.g., PSFs.  
We will address this issue in a future paper (Dahlen et al., in
preparation). 

The faint-end upturn suggests that a single Schechter function does not
provide a good representation of the early-type LF. To investigate how the
upturn affects results, we also fit the Schechter parameters after
excluding the faint population, similar to the approach in Madgwick et al. 
(2002) and Wolf et al. (2003). Results are given in Tables $1-3$.
As expected, we find that the faint-end slope is less steep after excluding
the upturn. The effect is, however, small, mainly because the relative
errors at the faint-end are large, giving low weight to these points 
when deriving the parameters over the full LF. 

In case of the $J$-band, we note that due to the upturn, it is possible to 
fit the LF with a straight line, corresponding to a pure power law.  
This is why there is no upper boundary on the early-type characteristic 
magnitude in the right panel of Figure \ref{fig4}.

\subsubsection{Luminosity Function Evolution}
\begin{figure*}
\centerline{
\psfig{figure=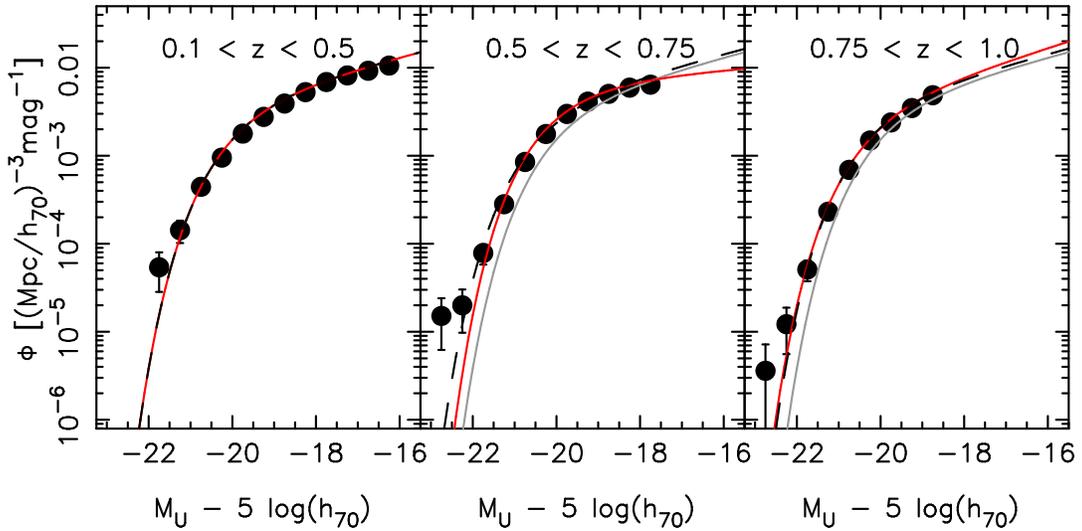,height=7.0cm}}
\caption{Rest-frame $U$-band luminosity function for redshift
  intervals $0.1<z<0.5$~(left panel), $0.5<z<0.75$~(mid panel), and
  $0.75<z<1.0$~(right panel). Solid red lines show the best fit Schechter
  function, while dashed (black) lines show the best-fitting Schechter
  function where the faint-end slope, $\alpha$, is fixed to the value
  measured in the lowest redshift bin. For comparison, we show with gray line 
  in the mid and high redshift bins the best-fitting Schechter function 
  found in the low redshift bin. 
\label{fig5}}
\end{figure*}

\begin{figure*}
\centerline{
\psfig{figure=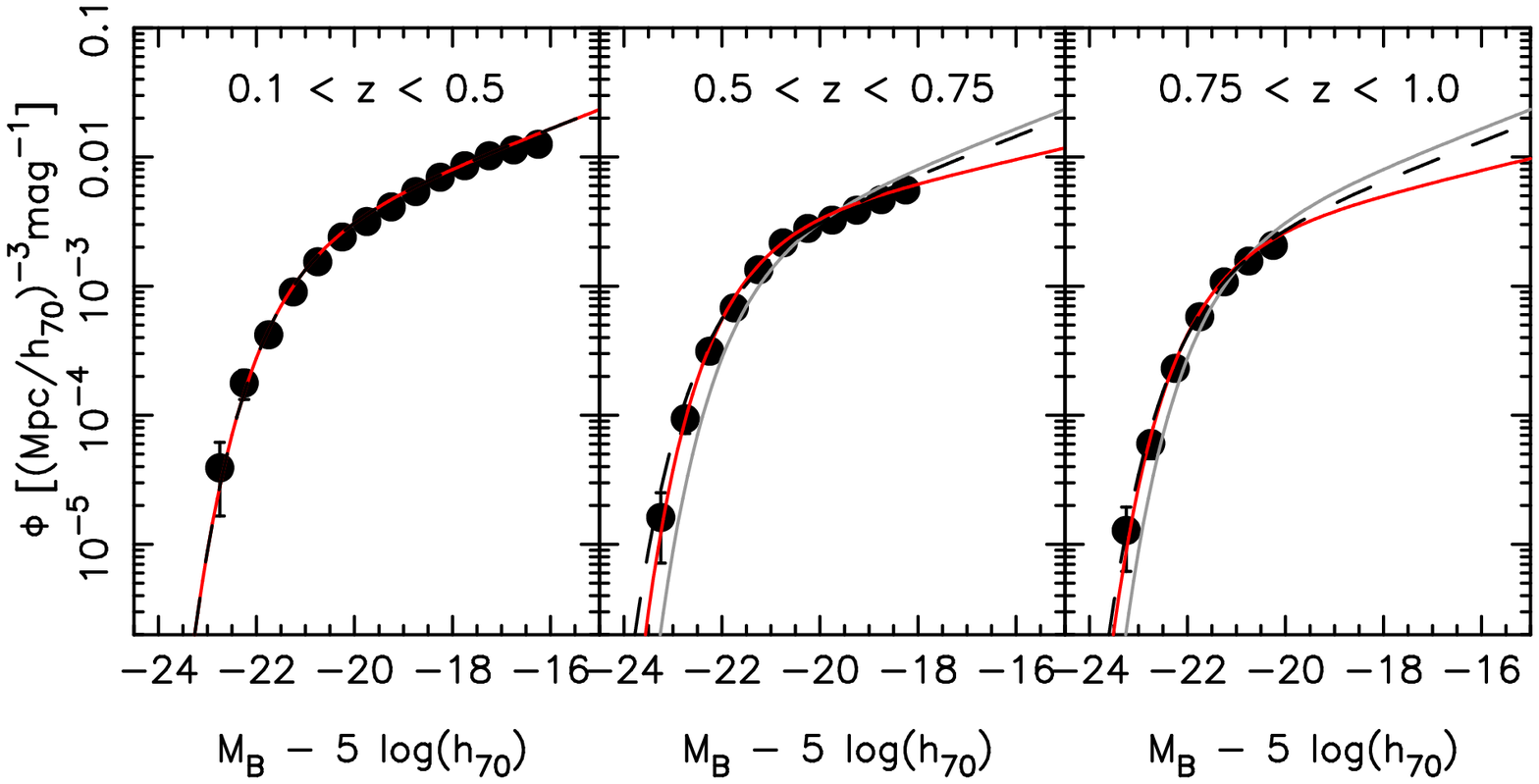,height=7.0cm}}
\caption{Rest-frame $B$-band luminosity function. Symbols are the same 
  as in Figure \ref{fig5}.
\label{fig6}}
\end{figure*}

\begin{figure*}
\centerline{
\psfig{figure=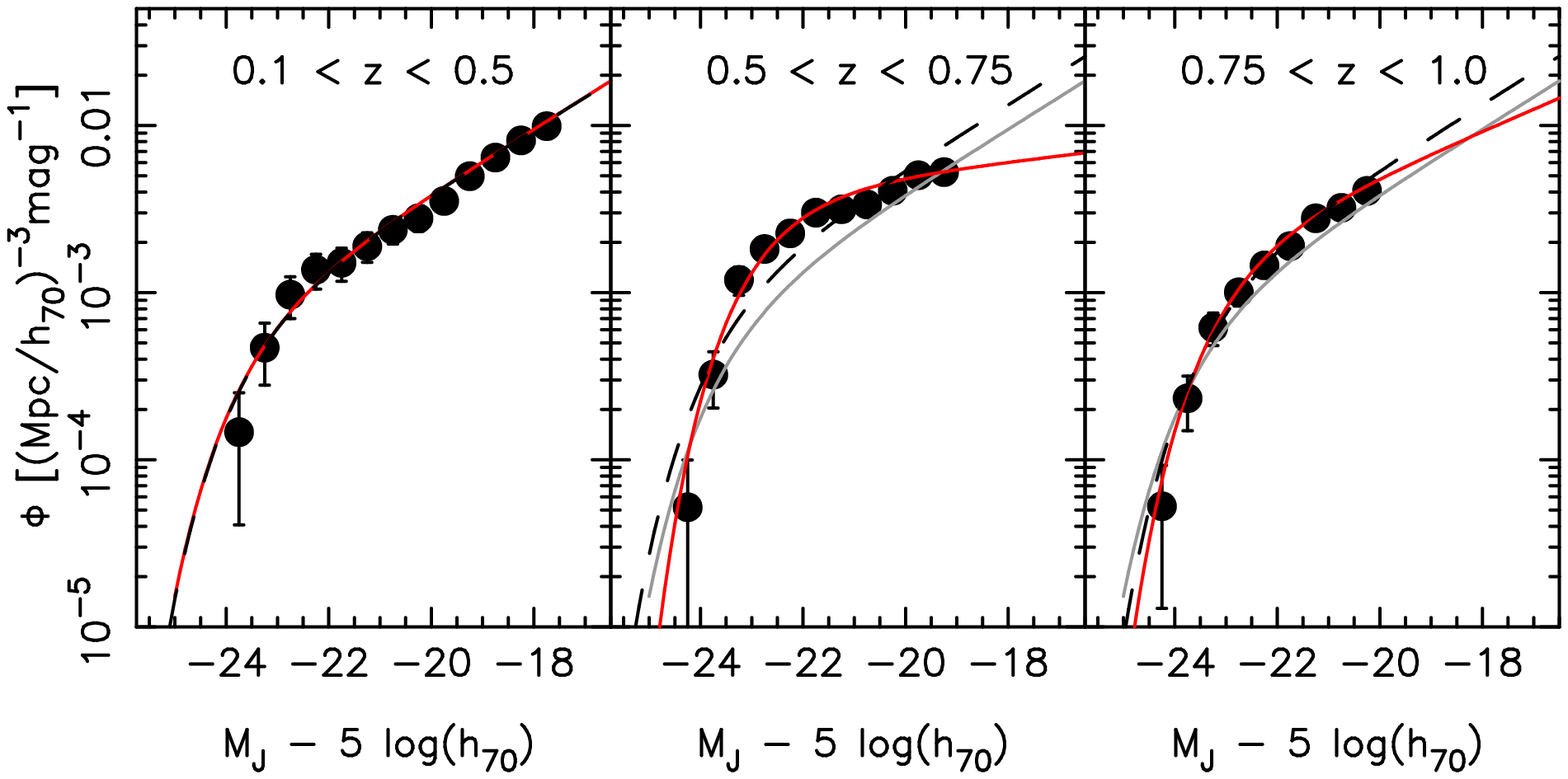,height=7.0cm}}
\caption{Rest-frame $J$-band luminosity function. Symbols are the same
  as in Figure \ref{fig5}.
\label{fig7}}
\end{figure*}

In Figures \ref{fig5} $-$ \ref{fig7}, we show the composite rest-frame
WFI-based $U-$, $B-$~and ISAAC-based $J$-band LFs in three redshift
bins $0.1<z<0.5$,~$0.5<z<0.75$~and $0.75<z<1$.
Because we do not reach the same depth at
all redshifts, we have to be careful when studying changes in
$M^*$~and $\alpha$ with redshift; i.e., in order to determine
$\alpha$, it is desirable to reach at least $\sim3$~mag fainter than
$M^*$. Also, due to the coupling between $M^*$~and $\alpha$~in
equation (4), the determination of $M^*$~is affected by the depth of the
survey.

To address this, we use the following two approaches.  First, we
fit both $M^*$~and $\alpha$~in each redshift interval, using different
limiting absolute magnitudes determined from the completeness in each
bin. The resulting best-fitting Schechter functions are shown as solid
red lines in Figures \ref{fig5}$-$\ref{fig7}.  Second, we fix the
faint-end slope to that found in the `quasi-local' sample
($0.1<z<0.5$~bin), and calculate the characteristic magnitude and
normalization in other redshift intervals. The resulting Schechter
functions are shown as dotted lines in Figures
\ref{fig5}$-$\ref{fig7}.  For comparison, we show in the medium and
high redshift bins the Schechter fit derived in the low redshift bin
as gray lines.  Results from the fits are also listed in Tables $2-4$.

The characteristic magnitudes in both the $U-$~and $B$-band LFs become
brighter by $\sim0.3$~mag between \~{z}=0.4 and \~{z}=0.9. For
the faint-end slope and the normalization, there is no clear trend
with redshift. In the $J$-band, there is an opposite trend in the
characteristic magnitude, i.e., it becomes somewhat fainter at higher
redshifts. Again, there is no clear trend in the other parameters over
this redshift range. The faint-end slope in all three bands is
consistent with a value $\alpha\sim-1.3$~- $-1.4$.

Due to the coupling between $M^*$~and $\alpha$~in the Schechter
function, their evolution with redshift is not independent and a trend
in one parameter may be correlated with the other. A better way to
characterize the evolution of the LF, as well as the contribution to
the total LF from the type dependent LFs, is to integrate the LF over
all magnitudes to derive the average luminosity density.  In the next
Section we investigate this further.

To extend the study of the rest-frame $U$-, $B$- and $R$-band LFs to 
$z\sim2$, we use the $K_s$-selected sample. This is important since galaxies 
start to drop-out of the $R$-band selected surveys at $z\gsim1$, 
where this band probes wavelengths shorter than the 4000\AA-break. Therefore, 
using an optically selected sample beyond this redshift only detects galaxies
that are very blue. As an example, with our limiting magnitudes
$R<24.5$~and $K_s<23.2$, we are able to detect galaxies with
$M_B\lsim-17.4$~and $M_B\lsim-16.9$~at $z=0.5$~in the two bands,
respectively (assuming a Sbc galaxy template). At $z=1.5$, we can
detect galaxies with $M_B\lsim-21.9$~and $M_B\lsim-19.7$, using the
$B-$ and $K_s$-selected catalogs respectively, clearly showing that we
reach significantly fainter rest-frame $B$-band magnitudes at high redshift
using the $K_s$-selection. It is therefore useful to directly compare
the LF derived using exactly the same methods on the $R$-selected and
$K_s$-selected catalogs.

\begin{figure*}
\centerline{
\psfig{figure=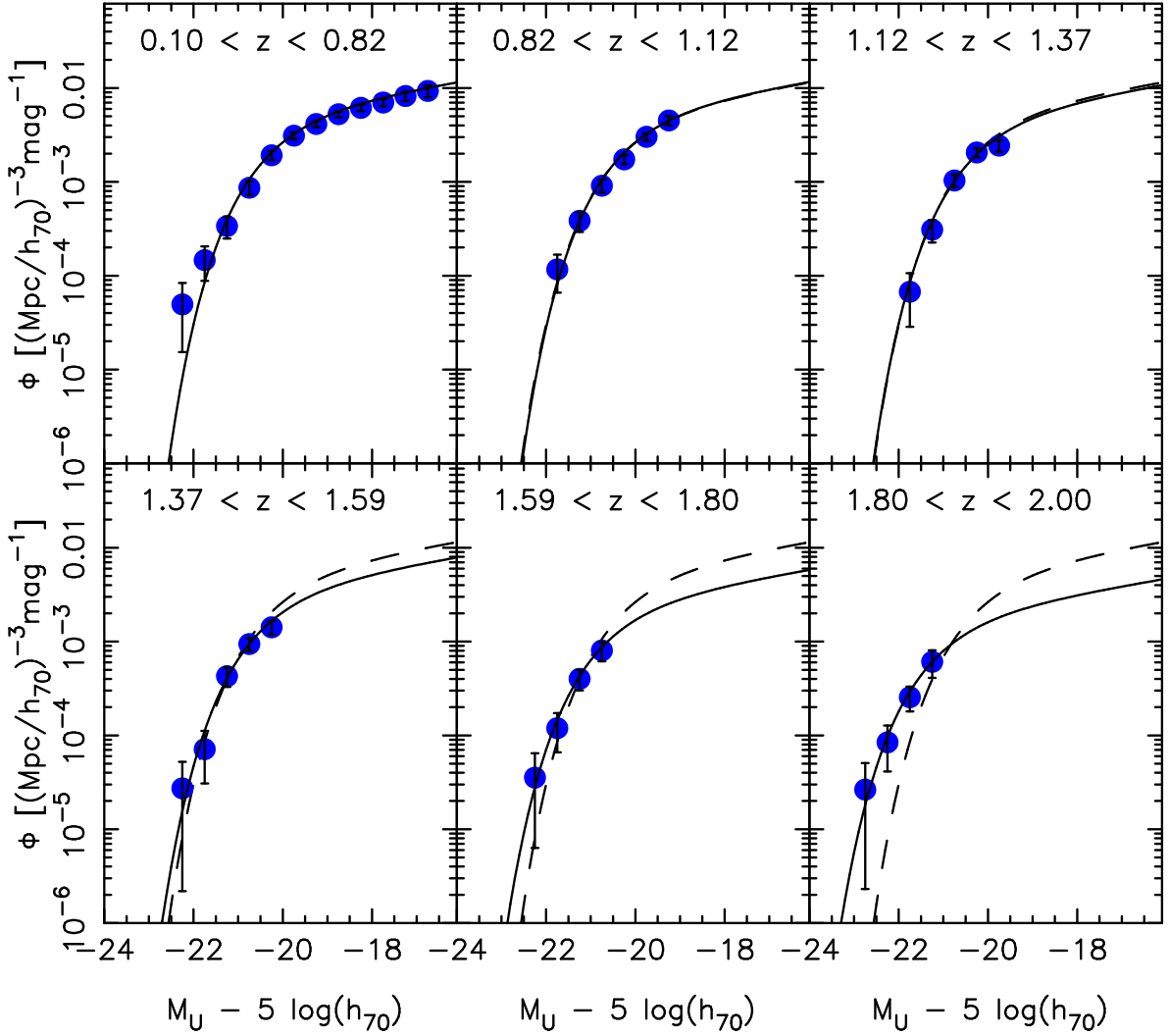,height=14.0cm}}
\caption{Rest-frame $U$-band luminosity function, based on the
  $K_s$-band selected catalog. Each redshift bin contains the same
  comoving volume. The solid line shows the best fit Schechter
  function where the faint-end slope, $\alpha$, has been fixed to the
  value measured in the lowest redshift bin.
\label{fig8}}
\end{figure*}

\begin{figure*}
\centerline{
\psfig{figure=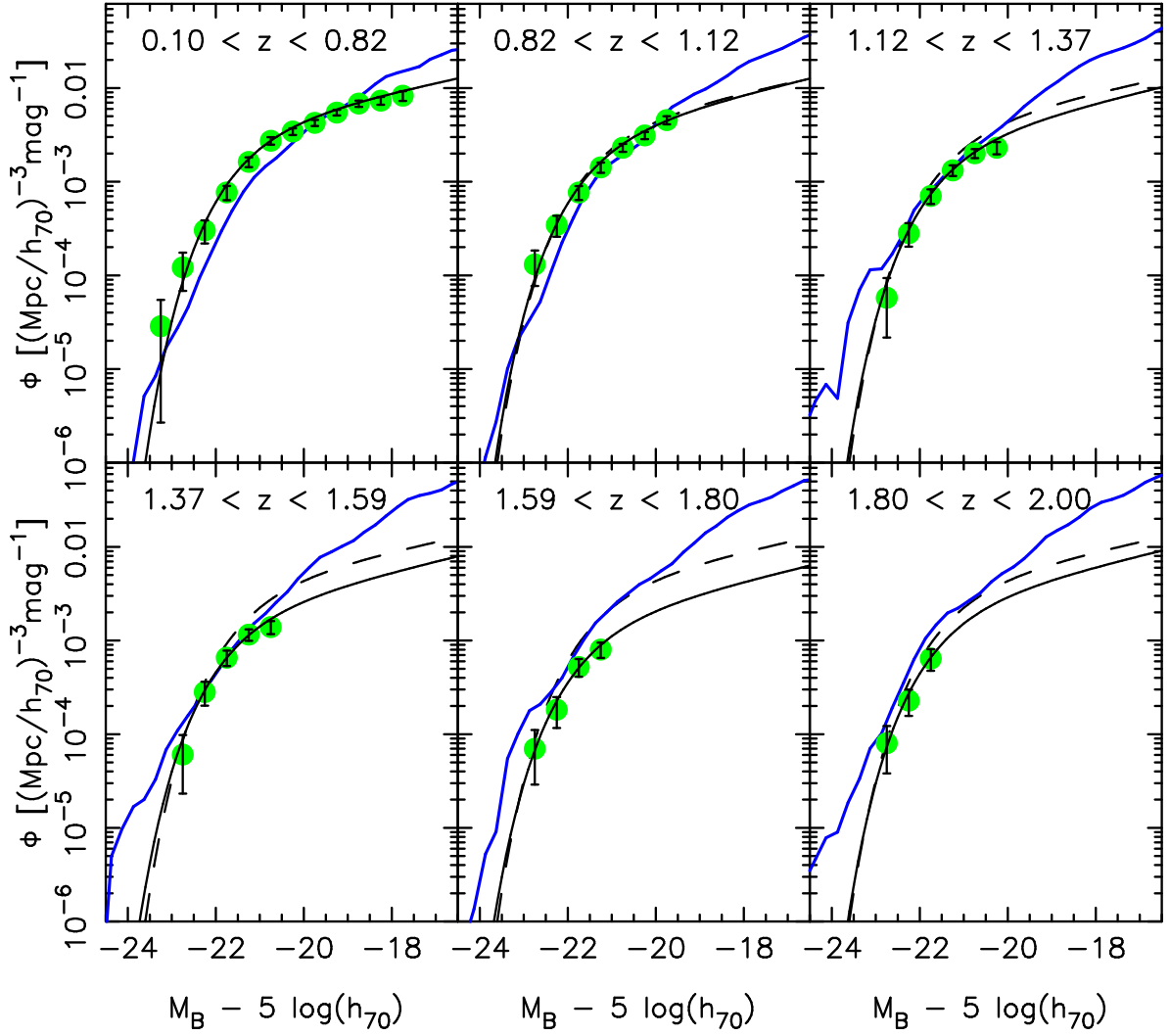,height=14.0cm}}
\caption{Rest-frame $B$-band luminosity function (for more information see
Figure \ref{fig8}). Blue lines show results from a semi-analytic model (see text). 
\label{fig9}}
\end{figure*}

\begin{figure*}
\centerline{
\psfig{figure=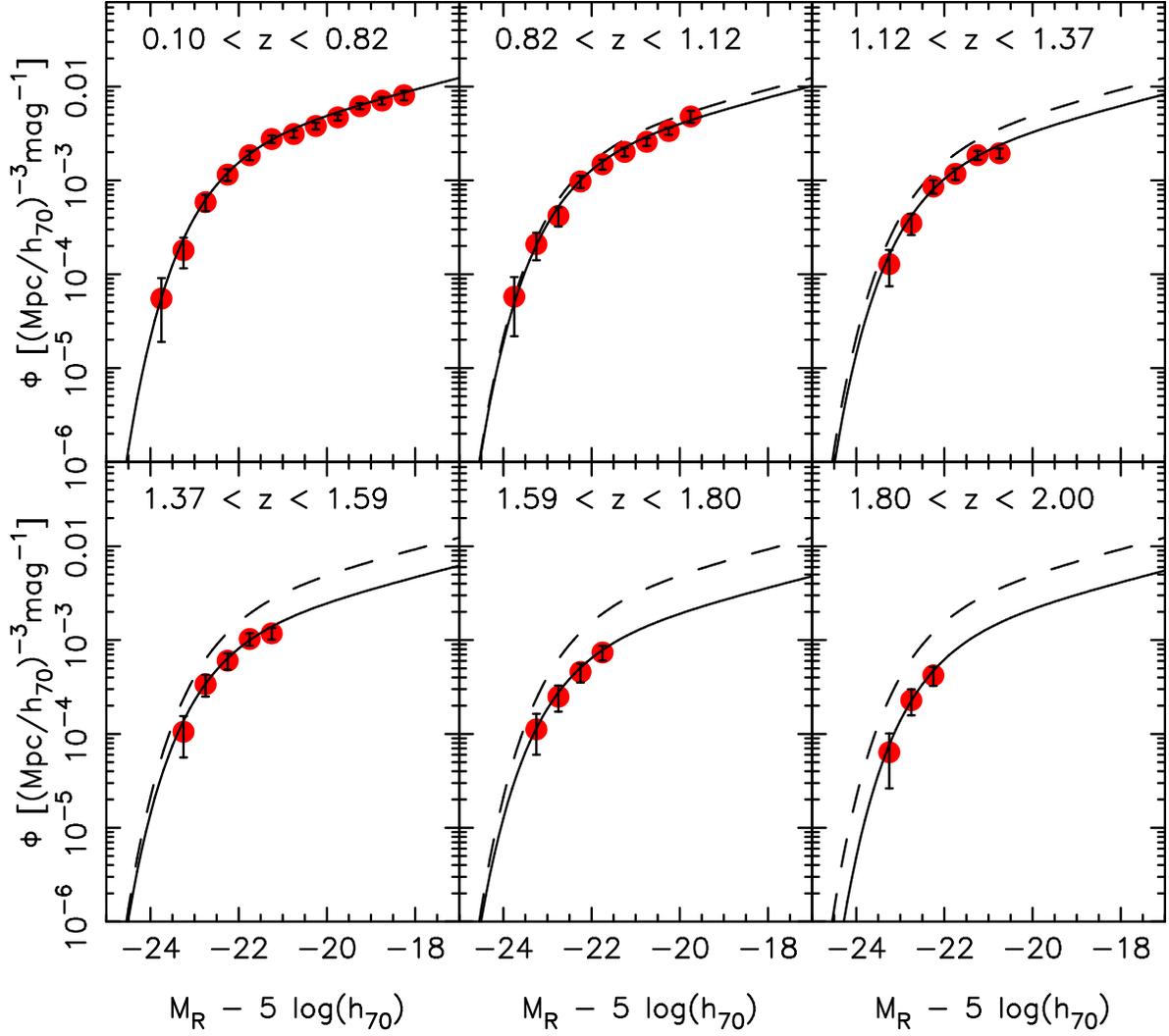,height=14.0cm}}
\caption{Rest-frame $R-$band luminosity function (for more information see
Figure \ref{fig8}).
\label{fig10}}
\end{figure*}

In Figures \ref{fig8}-\ref{fig10}, we show the evolution of the 
rest-frame $U$-, $B$-, and $R$-band LF in six roughly equal comoving volume 
redshift bins from $z=0.1$ to $z=2$. In the lowest redshift bin, we fit 
$M^*, \alpha$, and $\phi^*$, while for the higher redshift bins we adopt the 
faint-end slope from the lowest redshift bin and fit only $M^*$ and $\phi^*$.
The resulting best-fit Schechter functions are shown in the Figures as solid
lines and are listed in Table \ref{table4}. For reference, we show in
the five highest redshift bins the best fitting Schechter function
derived in the low redshift bin with dashed lines.  

Examining Figure \ref{fig8} and Table \ref{table4}, we find that the 
characteristic magnitude in the $U$-band brightens by $\sim0.9$ mag from 
{\it \~{z}}$\sim$0.6 to {\it \~{z}}$\sim$1.9, which is a similar evolution 
with redshift as found in the $R$-selected sample. The $U$-band also shows a 
strong decline in $\phi^*$~at higher redshift. We see no strong evolution 
in $M^*$~with redshift in rest-frame $B$- and $R$-bands, but note a decline 
in $\phi^*$~at higher redshift. 

We find good agreement between the $R$-selected and $K_s$-selected luminosity 
functions in the redshift range where they overlap. E.g., at {\it \~{z}}=0.64 
($R$-selected) and {\it \~{z}}=0.62 ($K_s$-selected), we find $M^*_B=-21.46$~
and $M^*_B=-21.43$, as well as $M^*_U=-20.08$~and $M^*_U=-20.28$, for the two 
different samples, respectively.

In Figure \ref{fig9} we compare our results with the predictions of a
semi-analytic model of galaxy formation that has previously been shown
to produce good agreement with the optical luminosity function at
redshift zero and the rest-frame $UV$~LF of Lyman break galaxies at
$z=3$ (Somerville \& Primack 1999; Somerville et al.  2001,
hereafter SPF2001). The predictions are based on the `collisional
starburst model' described in SPF2001 and computed using the same
cosmology used throughout this paper ($\Omega_M$=0.3,
$\Omega_{\Lambda}$=0.7, $H_0$=70 km s$^{-1}$~Mpc$^{-1}$).

We find qualitatively good agreement between the observations and the model
predictions at all redshifts to the completeness limit in each bin. At
magnitudes fainter than we can observe, there is a discrepancy between
the models and the extension of the Schechter function fits in the
sense that the model systematically produces more faint galaxies than
the Schechter function would imply.  This discrepancy increases with
redshift and will affect the comparison between the observed and
predicted evolution of the luminosity density as discussed below. It
is not yet clear whether the models over-predict the number of faint
galaxies at high redshifts, or whether our assumption of a fixed,
non-evolving, faint-end slope is incorrect. This demonstrates that
probing to fainter magnitudes would provide a strong test of the
models.

In the discussion in \S \ref{sec:discussion} we compare our results on
the LF with results taken from the literature.
\subsection{Luminosity Density}
\label{sec:results:lumdens} 

\begin{figure*}
\centerline{
\psfig{figure=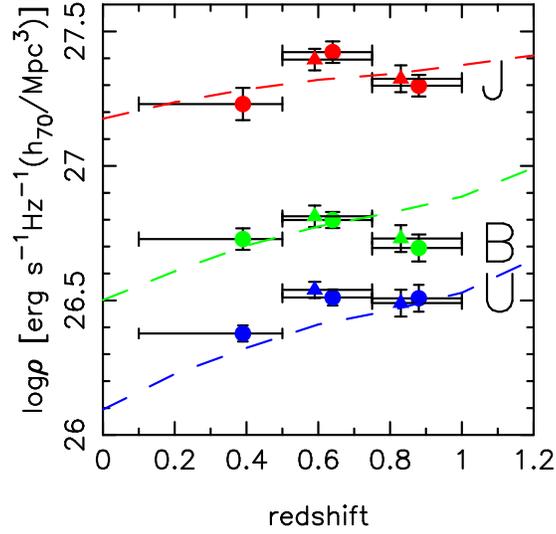,height=7.0cm}}
\caption{The evolution of the luminosity density as a function of
  redshift for rest-frame $U$-, $B$-, and $J$-bands is shown as filled
  circles. The luminosity densities in the two higher redshift bins
  assuming a fixed faint-end slope are shown with triangles (circles
  and triangles overlap in the lowest redshift bin). Predictions from
  a semi-analytical model are shown as dashed lines.
\label{fig11}}
\end{figure*}

The luminosity density is calculated by integrating the LF over
luminosity. To correct for incompleteness, we approximate the LF by
the estimated Schechter function parameters and integrate this over
all luminosities. The luminosity density is then given by:
\begin{equation}
\rho_{\nu} = \int L_{\nu}\phi(L_{\nu},z)dL_{\nu}=\Gamma(2+\alpha)\phi^*L_*.
\end{equation}
In Figure \ref{fig11}, we show the evolution of the luminosity density
with redshift over the $0.1<z<1$ range for rest-frame $U-$, $B-$
(WFI-based) and $J$-bands (ISAAC-based).  Filled circles show the
evolution when fitting all Schechter parameters in each bin ($M^*$,
$\alpha$, and $\phi^*$) while open triangles show the evolution
assuming a fixed faint-end slope. The latter are somewhat offset in
the x-direction for clarity.

We find that between {\it \~{z}}$\sim$0.4 and {\it \~{z}}$\sim$0.9, there is 
a mild increase in the $U$-band luminosity density by a factor $\sim1.3$. In
the $B-$~and $J$-bands, there is an increase in the luminosity density
between the low and mid redshift bins and a slight decrease at higher
redshifts. The overall trend in the $B-$~and
$J$-band luminosity density is consistent with being roughly constant
over this epoch, especially when cosmic variance is considered.
We obtain similar results, well within the $1\sigma$-errors, whether
we leave the faint end slope free or fix it to the
value in the lowest redshift bin.

In Figure \ref{fig11}, we also plot predictions from semi-analytical models
(SPF2001) as dashed lines. The models are in broad agreement with 
observations predicting a stronger evolution in the $U$-band compared to 
longer wavelength bands, especially the $J$-band. The observed 'bump' in
the luminosity densities in the mid redshift bin, which is contrary to the 
more smooth model prediction, suggests that there is an over abundance in the
$0.5<z<0.75$~bin compared to the $0.75<z<1.0$~bin due to cosmic variance.
An over abundance in this redshift range in the CDFS is also found by the 
VVDS (Le F\`{e}vre et al. 2004), who suggests that there is a wall-like 
pattern at $z\sim0.7$.

\begin{figure*}
\centerline{
\psfig{figure=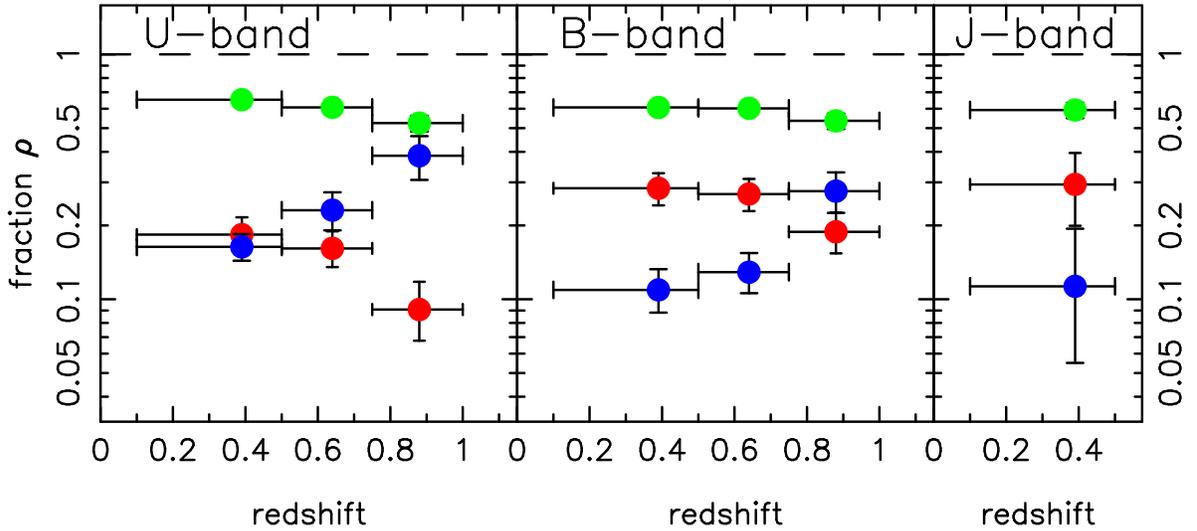,height=7.0cm}}
\caption{Fractional contribution from early-type galaxies (red
  points), late-type galaxies (green points), and starbursts (blue
  points) to the total luminosity density in $U$-band, $B$-band and
  $J$-band. For the $J$-band, we only show results in the low redshift bin
  where statistics are sufficient for a determination of the fractions. 
\label{fig12}}
\end{figure*}

In Figure \ref{fig12}, we show the fractional contribution to the
luminosity density at $0.1<z<1.0$ from galaxies of different
spectral types. In the $U$- and $B$-band, we show this for all three
redshift bins, while for the $J$-band we plot only the lowest redshift
bin.  In the higher redshift bins, the small numbers of galaxies in the
$K_s$-selected catalog does not allow us to calculate fractional
contributions with sufficient accuracy. 
Note that the redshift evolution in the fractions should only be 
marginally affected by cosmic variance since the total number of galaxies do 
not enter explicitly here. In the lowest redshift bin, we
note that a large part of the luminosity density comes from late-type
galaxies in all bands with the main difference being that the
starburst contribution is higher and the early-type contribution is
lower in the $U$-band compared to the other bands. In both the $B-$
and $J$-bands, about 30\% of the light comes from early-type galaxies.

While there is only a weak trend in the evolution of the composite
luminosity density in the $U$- and $B$-band (Figure \ref{fig11}), there
is significant evolution of type-specific luminosity densities, as
shown in Figure \ref{fig12}. In the $U$-band, the fraction of the
luminosity density contributed by the starburst types increases by a
factor $\sim2.4$~over the redshift range, while the fractional
contribution of both late-type and early-type galaxies decreases with
redshift.  In absolute terms, the starburst luminosity density
increases by a factor $\sim3$~over the redshift interval considered. The
overall increase in the $U$-band luminosity density is therefore
mainly due to the increase in the contribution by the starburst
population. The same trends seen in the $U$-band are also present in
the $B$-band.  However, while the starburst fraction increases, as in
the $U$-band (here a factor $\sim2.5$), because the overall fraction
of light contributed by starburst types is lower, this does not affect
the evolution of the total luminosity density as strongly as in the
$U$-band.

\begin{figure*}
\centerline{
\psfig{figure=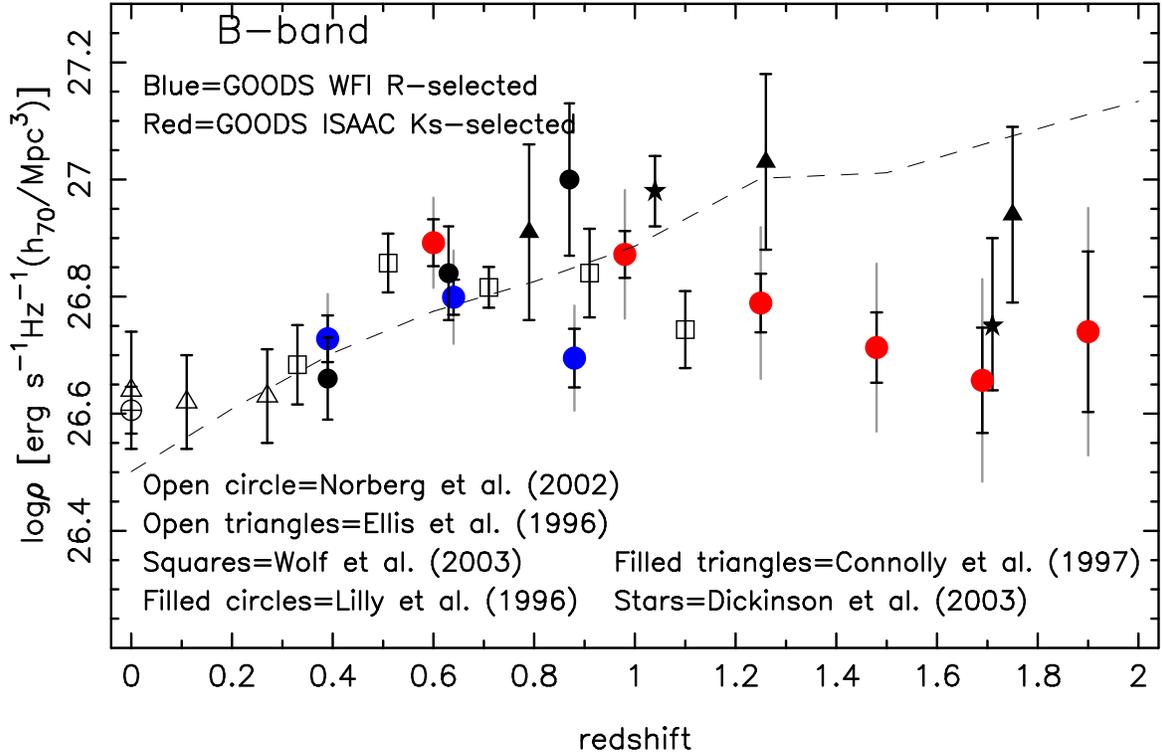,height=10.0cm}}
\caption{Evolution of the rest-frame $B$-band luminosity density to
  $z\sim2$. Blue dots show the GOODS measurements based on the
  $R$-selected sample, while red dots show results derived from the
  $K_s$-selected sample. Black error-bars represent statistical errors
  only, while gray error-bars also include cosmic variance.  Results
  from the literature are taken from the 2dF-GRS (Norberg et al. 2002;
  open circle), Autofib Redshift Survey (Ellis et al. 1996; open
  triangles), Canada-France Redshift Survey (Lilly et al. 1996; filled
  circles), COMBO-17 (Wolf et al. 2003), the original HDF (Connolly et
  al. 1997; filled triangles), and the HDF+NICMOS (Dickinson et
  al. 2003; stars). Dashed line shows the prediction from a semi-analytical
  model, as described in the text.
\label{fig13}}
\end{figure*}

In Figure \ref{fig13}, we show a compilation of results from the
literature for the rest frame $B$-band luminosity density from $0\la z
\la 2$, along with our results based on the $R$-selected and
$K_s$-selected catalogs. We show error-bars corresponding to the Poisson
errors only (black error-bars), and error-bars including cosmic variance
(grey error-bars), estimated as described in
\S\ref{sec:methods:cosvar}.  We find that the luminosity densities
derived from the $R-$~and $K_s$-selected samples show good
agreement in the redshift range where they overlap.

We also find consistency between our results and those from literature,
supporting a mild increase in the $B$-band luminosity between the
local value and redshift $z\sim1$. In the overlapping redshift range
(0.3\lsim $z$~\lsim 1.2), there is also a good agreement between our
results and COMBO-17 (Wolf et al. 2003). Compared to the local value
from Norberg et al., both this investigation and COMBO-17 find an
increase in the rest-B luminosity density by a factor $\sim1.7$
between $z\sim0$~and $z\sim0.9$. Over the same redshift range, the
Lilly et al. (1996) results, when converted to our assumed cosmology, imply a
stronger increase (factor $\sim2.5$), but the difference is within the
errors. Note that the original increase reported by Lilly et al. was
steeper (up by a factor of $\sim3.7$~compared to the local value from
Norberg et al. 2002) due to their use of a cosmology with $q_0=0.5$~and
$\Omega_0=1$.  At high redshifts ($z>1.2$), we find somewhat lower
luminosity densities compared to Connolly et al. (1997), however, we
are in good agreement with the data point at $z\sim$1.7 from Dickinson
et al. (2003).  Note here that both the Connolly et al. and the Dickinson et
al. results are based on data from the Hubble Deep Field, but Dickinson et al.
include NICMOS NIR data, analogous to the inclusion of ISAAC
NIR data in our investigation, so likely resulting in more
reliable photometric redshift estimates.  Over the redshift range
$0.5\lsim z \lsim 2 $, all the data are consistent with a constant
luminosity density in the $B$-band.

Also shown in Fig.~\ref{fig13} is a comparison with the semi-analytic
model predictions. The models predict a continuous decline in the
luminosity density from $z\sim2$ to the present. As seen in
Fig.~\ref{fig9}, the models agree well with the observed luminosity
functions over the magnitude range directly probed by the
observations. This implies that the stronger evolution in the
luminosity density predicted by the models is entirely due to a
population of galaxies too faint to be directly constrained by our
observations. Whether or not this population indeed exists remains to
be seen.

\section{Discussion and Conclusions}
\label{sec:discussion}
\subsection{Optical bands}
\label{sec:discussion:optical} 
Using the GOODS dataset, including both \HST~ACS and ground-based
photometry from $U$ through $K_s$, we have used photometric redshifts to
study the evolution of the LF and luminosity density over the redshift
range $0\la z \la 2$. We have derived the rest frame optical ($U-$~and
$B$-band) LF from $0.1 \la z \la 1$ based on an $R$-selected catalog,
and used a smaller area but very deep $K_s$-selected catalog to derive
the rest $U$-, $B$- and $R$-band LF to $z\sim2$ and the NIR ($J$-band) LF to
$z\sim1$.

At $z\la1$, we detect a brightening of the characteristic magnitude
with increasing redshift in optical bands based on the $R$-band selected 
sample. A brightening in the $U$-band is also found to $z\sim$2 using the 
$K_s$-selected sample, while there is no clear trend in the $B$- or the 
$R$-bands at $z\le0.7$, based on the latter sample.

\begin{figure*}
\centerline{
\psfig{figure=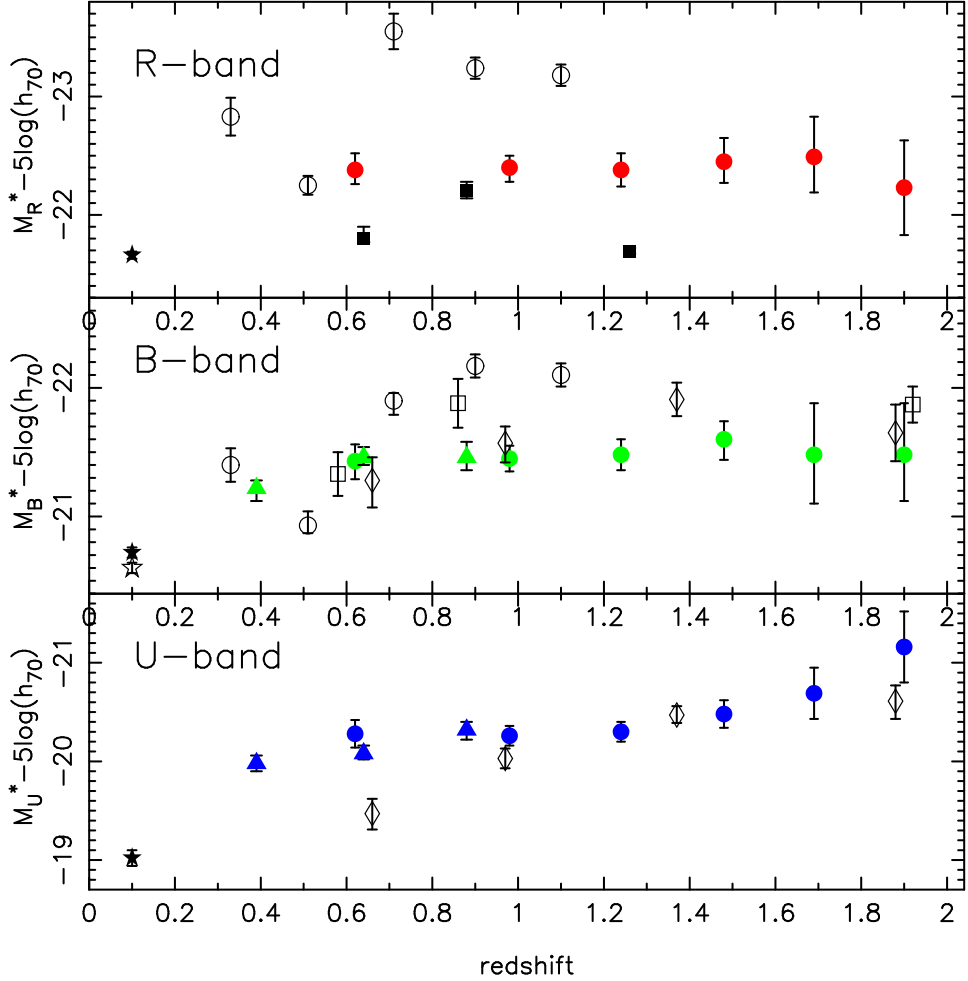,height=13.0cm}}
\caption{Comparison between characteristic magnitudes in this work and values 
taken from literature. GOODS measurements are shown as triangles ($R$-band 
selected sample) and filled circles ($K_s$-selected sample), with blue, green 
and red colors showing rest-frame $U$-, $B$-~and $R$-band, respectively. 
Literature measurements are taken from SDSS (Blanton et al. 2001; filled 
stars), 2dF-GRS (Madgwick et al. 2002; open star), FORS Deep Field (Gabasch 
et al. 2004; open diamonds), COMBO-17 (Wolf et al. 2003; open circles), Poli 
et al. (2003; open squares) and Chen et al. (2003; filled squares).
\label{fig14}}
\end{figure*}

In Figure \ref{fig14} we compare the characteristic magnitude derived 
from GOODS with results taken from the literature. The GOODS data are shown 
as circles ($K_s$-selected) and triangles ($R$-selected), with blue, green 
and red colors representing rest-frame $U$-, $B$- and $R$-band, respectively. 
Results from literature are converted to our adopted cosmology and AB 
magnitudes when necessary.

In the $U$-band (bottom panel) there is an excellent agreement between our 
results and the results from the FORS Deep Field (Gabasch et al. 2004; open 
diamonds). There is a clear trend of a brightening of $M^*$~with redshift. 
Compared to the 'local' value from SDSS (Blanton et al. 2001), we find a 
brightening of the characteristic magnitude by $\sim$2.1 mag to $z\sim1.9$.
This is consistent with recent results from VVDS (Ilbert et al. 2004), who
measures a brightening of $M^*_U$~by 1.8-2.4 magnitudes between $z=0.05$~and
$z=2.0$.
In the $B$-band (mid panel) there is also good agreement between GOODS data 
and the FORS Deep Field, as well as with results from Poli et al. (2003; 
open squares). The results from COMBO-17 (Wolf et al. 2003; open circles) are 
consistent with the other results, but with higher scatter and
with the two highest redshift points significantly brighter than the other 
measurements. Note, however, that one reason for the scatter in the Wolf et 
al. points is due to the use of a non-fixed faint-end slope when determining 
$M^*$ in COMBO-17. The covariance between $M^*$~and $\alpha$ causes a higher 
scatter in $M^*$~between bins compared to the case where $\alpha$~is fixed to
a common value in all bins. The brightening of $M^*_B$ between the local 
value from SDSS and our measurement at $z\sim1.9$~is $\sim 0.8$~mag. The 
evolution is strongest at $z\la 0.7$. At higher redshifts, our results are 
consistent with a flat evolution.
Finally, in the $R$-band (top panel), the scatter between measurements is 
the largest.
There is no clear trend in either GOODS, COMBO-17 or data from Chen et al. 
(2003; open triangles) of a brightening of $M^*$~with redshift. When comparing
to the local SDSS measurement, both GOODS and COMBO-17 find brighter 
characteristic magnitudes. The evolution in the GOODS data is $\sim0.7$~mag
to $z\sim1.9$, with to strongest evolution at $z\la 0.7$, similar to the
other bands. 
 
In summary, the non-local measurements show characteristic magnitudes
that are brighter than the low redshift ('local') values taken from SDSS 
(Blanton et al. 2001; filled star) and 2dF-GRS (Madgwick et al. 2002; open 
star), with a more significant difference in shorter wave-length bands.

The faint-end slope of the LF in all bands is consistent with
$\alpha\sim-1.3$~- $-1.4$, and we see no evidence for strong evolution in 
the faint-end slope to $z\sim1$ in any band.

We find that the shape of the LF is strongly dependent on spectral
type, consistent with the results from 2dF-GRS (Madgwick et al. 2002) and
COMBO-17 (Wolf et al. 2003). In particular, we note that the
starburst-type population has a characteristic luminosity that is
significantly fainter than the composite LF, and that the early-type
LF has more of a Gaussian shape, with a possible upturn at faint
magnitudes.  This upturn in the early-type LF has also been reported
by Madgwick et al. (2002) based on the 2dF-GRS. We
also find significant evolution with redshift in the relative
contribution from different types. Most prominently, the contribution
from starburst galaxies to the total luminosity density increases with
redshift. This is most evident in the $U$-band, where the starburst
fraction increases by a factor $\sim3$~over the redshift range
measured. At the same time, we note a significant decrease in the
contribution from early-type galaxies with increasing redshift. Again,
these results are consistent with similar findings based on COMBO-17
reported in Wolf et al. (2003).

The LFs derived from our $R$-selected and deep $K_s$-selected catalogs
agree well out to $z\sim1$, indicating that at $z \la 1$, $R$-selected
surveys ($R\la 24$) are not likely to be missing a large population of
the objects that would be selected in the $K$-band at $K\la22$. This
is encouraging for the large upcoming $R\la24$ selected spectroscopic
surveys such as DEEP and VVDS.

Integrating our LF results using equation (8), we obtain estimates of 
the luminosity density in the rest-frame $U$-, $B$-, and $R$-bands to 
$z\sim2$ and $J$-band to $z\sim1$. The $U$-band shows a
rather mild increase (about a factor of 1.3) between {\it \~{z}}$\sim$0.4 
to {\it \~{z}}$\sim$0.9, based on the $R$-selected sample. At $z\ga 1$, we
do not detect any clear trend using the $K_s$-selected sample.  The $B$- 
and $R$-band luminosity densities are consistent with being constant to 
$z\sim2$. This evolution could be significantly
underestimated, however, if the faint end slope at high redshift is
actually steeper than we have assumed.

We find that our luminosity function determinations agree fairly well with the 
predictions of a semi-analytic model of galaxy formation over the magnitude 
and redshift range probed by the observations. The shape of the predicted 
luminosity function deviates from a Schechter form, especially at high 
redshift, producing an excess of both $L>L_{*}$ and $L<L_*$ galaxies. These 
problems are endemic to CDM-based models and are well known. It is clear, 
though, that the models produce at least as many luminous ($L>L_*$) galaxies 
at high redshift ($1 \la z \la 2$) as are implied by our observations, while 
some previous works have suggested that semi-analytic models may have 
difficulty in this regard (e.g. Somerville et al. 2004b; 
Glazebrook et al. 2004 Nature, 430, 181). Because of the 
combined effects of a decrease in the excess of bright galaxies relative to a 
Schechter fit, a mild decrease in $L_*$, and a flattening of the faint end of 
the LF with time, the models predict a monotonically decreasing luminosity 
density from $z\sim2$ to the present, in contrast with the rather flat 
luminosity density implied by our observations. However, the models do 
predict a much flatter dependence of luminosity density on time in the NIR 
bands than in the optical and UV, as seen in the data, consistent with an 
overall decrease in the global "specific star formation rate" (star 
formation rate per unit stellar mass) over time.

\subsection{NIR bands}
\label{sec:discussion:nir} 
In the $J$-band, we find a trend where the characteristic magnitude, 
$M_J^*$, gets fainter over the range {\it \~{z}}$\sim$0.4 to {\it \~{z}}$\sim$
0.9 by $\Delta M_J^*\sim0.6$. If we fix the faint-end slope to the value in 
the lowest redshift bin, the evolution is less significant, 
$\Delta M_J^*\sim0.3$. A mild fading of $M_J^*$ is also found by Pozzetti 
et al. (2003), who measures $\Delta M_J^*\sim0.14$~between redshifts 
{\it \~{z}}$\sim$0.5 to {\it \~{z}}$\sim$1.05. However, the error-bars are 
larger than this difference and results are therefore not significant. An 
opposite trend is found by Feulner et al. (2003), who find a brightening 
$\Delta M_J^*\sim-0.6$~between redshifts {\it \~{z}}$\sim$0.24 to 
{\it \~{z}}$\sim$0.48.

\begin{figure*}
\centerline{
\psfig{figure=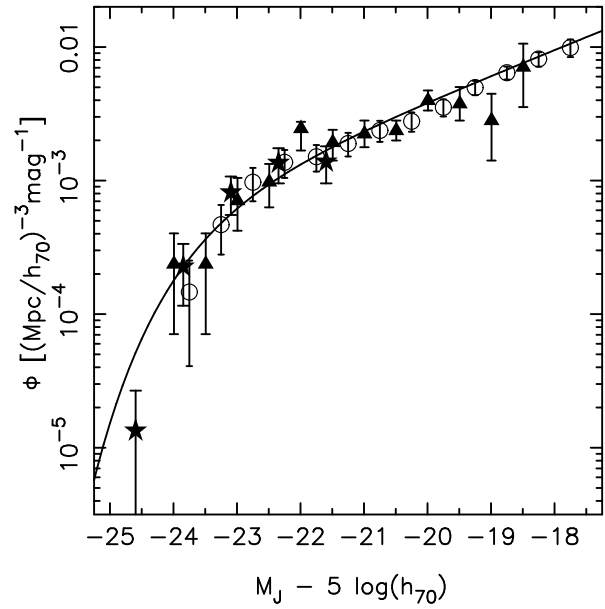,height=8.0cm}}
\caption{Rest-frame $J-$band LFs from this paper (open circles, {\it \~{z}}=0.39), Feulner et al. (2003, stars, {\it \~{z}}=0.48)
and Pozzetti et al. (2003, triangles, {\it \~{z}}=0.5 ).
\label{fig15}}
\end{figure*}

In Figure \ref{fig15}, we compare the $J$-band LF derived in this paper 
with the LFs from Pozzetti et al. (2003) and Feulner et al. (2003). We have 
chosen bins at similar redshifts for this comparison, i.e., {\it \~{z}}=0.39,
{\it \~{z}}=0.50 and {\it \~{z}}=0.88 for the GOODS, Pozzetti et al. 
and Feulner et al. (2003) data, respectively. Inspecting the Figure, we find an 
excellent agreement between the surveys. Despite this, when comparing at the 
Schechter function parameters derived in the different surveys, we find that 
numbers differ significantly. The characteristic 
magnitude is $M^*_J=$--23.68, --22.93 and --22.98, for this investigation, 
Feulner et al. (2003) and Pozzetti et al. (2003), respectively. The faint-end 
slope is $\alpha=$--1.48, --1.00 and --1.22, and the normalization is 
$\phi=$0.0008, 0.0026 and 0.0020 mag$^{-1}h_{70}^3$Mpc$^{-3}$, for the three 
investigations. However, comparing Schechter function parameters one-to-one 
can be misleading since there is a coupling between the different parameters.
As an alternative, we use equation (8) the derive the luminosity density in 
the three surveys. We find $log(\rho_{\nu})$=27.23, 27.23 and 27.20 
erg s$^{-1}$Hz$^{-1}$(h$_{70}$/Mpc$^3$) for the three surveys, respectively. 
This agreement is excellent and further stress that comparisons between only 
one of the Schechter parameters may be misleading. 
   
Figure \ref{fig15} also illustrates that the combination of depth and wide 
area in GOODS results in significantly better statistics, especially at the 
faint-end, compared to the other investigations.

We find a mild increase in the $J$-band luminosity density between 
{\it \~{z}}$\sim$0.4 and {\it \~{z}}$\sim$0.9, results are, however, 
also consistent with being constant.

At longer rest-frame wavelengths, Drory et al. (2003) and Caputi et al. (2004)
report a mild brightening of the characteristic magnitude with redshift
in the rest-frame $K$-band, opposite to the trend found here in the $J$-band. 
Deriving the redshift evolution in the rest-frame $K$-band 
using observed $K$-band relies on proper K-corrections. Local 
K-corrections in the NIR are well known, however, any redshift
dependence on the K-corrections will affect results.  
Also, a bias in the determination of $M_K^*$~may be introduced if 
there is a differential trend in the redshift evolution in $M^*$, i.e.,
the brightening gets larger at shorter wave-lengths, 
as reported by Ilbert et al. (2004) who find an evolution
in $M^*$~that is strongest in the $U$-band and becomes monotonically
weaker at longer rest-frame wavelengths (including $B$-, $V$-, $R$-,
and $I$-bands). As the 
observed $K$-band probes shorter rest-frame wavelengths at higher 
redshifts, the differential trend described above could mimic  
brightening in $M_K^*$~with redshift.

We also note that making the simple assumption that the evolution in
$M^*$~is proportional to wavelength, the results from Ilbert et
al. (2004) suggest that we might expect a `turn-over' at approximately
the NIR $J$-band, where $M^*$~starts to become fainter with
redshift, consistent with what is found here in the $J$-band.

Therefore, further investigations in the rest-frame NIR
are needed to firmly establish the evolution of the characteristic
luminosity. Observations in the mid-infrared with the Spitzer
Space Telescope will here be of great importance.

While the optical light (especially $U$-band) is related to the
underlying star-formation, the NIR (e.g., the $J$-band) light is
more directly related to the underlying stellar mass of the
galaxies. The opposite trends observed in the optical and NIR for
the characteristic magnitude therefore suggest a scenario in which the
star-formation rate in galaxies increases with redshift, while the
underlying stellar mass decreases; or equivalently in which the
specific star formation rate (star formation rate divided by stellar
mass) decreases with time. These trends are in qualitative agreement
with predictions from the hierarchical clustering scenario. In a
separate paper, we use the SEDs constructed here to estimate the
\emph{stellar mass} of each galaxy, and to directly estimate the rate
of stellar mass build-up in our sample over the redshift interval from
$z\sim2$ to the present (Somerville et al., in preparation).

 The steep
faint-end slope that we obtain in the $J$-band in our lowest redshift
bin ($\alpha=-1.48^{+0.06}_{-0.05}$) is consistent with the slope 
$\alpha=-1.22^{+0.22}_{-0.20}$~derived by Pozzetti et al. (2003). 
However, the faint-end slope we find seems inconsistent with the one
derived for nearby galaxies from 2MASS. E.g., Kochanek et al. (2001)
found $\alpha=-1.09\pm0.06$~in the 2MASS $K$-band. We would not
expect such a strong dependence on redshift or such a large difference
between the $J$- and $K$-bands. However, Kochanek et al. include 2MASS
galaxies with $M\lsim M^*+3$~mag when fitting the Schechter
parameters, while we reach significantly deeper, $M<M^*+6$~mag. To
investigate, we recalculated our Schechter parameters using a faint
cutoff at $M=M^*+3$~mag. We find $\alpha=-1.15^{+0.21}_{-0.20}$,
consistent with the results from 2MASS. The reason for the
significant increase in the steepness of the faint-end slope at the 
faint magnitudes reached in this investigation, is that we start to 
probe the abundant 
population faint starburst galaxies. This is evident in Figure \ref{fig4}, 
which shows that the steep starburst population makes the composite
LF turn steep at faint magnitudes. Contrary to this, the shallower 
2MASS do not reach this abundant population and therefore find a flatter
faint-end slope.
This illustrates how the depth
of the survey, together with the covariance between $M^*$~and
$\alpha$, can significantly affect the derived Schechter function
parameters. Moreover, another investigation based on NIR data that is
deeper than 2MASS but shallower than ours also found a steeper faint
end slope (Huang et al. 2003).


\acknowledgments{ We thank the GOODS team, in particular Rafal Idzi
  and Kyoungsoo Lee, for their efforts in data reduction and
  cataloging. Support for the GOODS \HST ~Treasury program was
  provided by NASA through grants HST-GO-09425.01-A and
  HST-GO-09583.01 from the Space Telescope Science Institute, which is
  operated by the Association of Universities for Research in
  Astronomy under NASA contract NAS5-26555. Based on observations collected 
  at the European Southern Observatory, Chile (ESO programmes 168.A-0485, 
  170.A-0788, 64.O-0643, 66.A-0572, 68.A-0544, 164.O-0561, 169.A-0725,
  267.A-5729 66.A-0451, 68.A-0375 164.O-0561, 267.A-5729, 169.A-0725, 
  and 64.O-0621). M.D. and L.A.M. acknowledge support from the Spitzer
  Legacy Science Program, provided by NASA through contract 1224666 issued
  by the Jet Propulsion Laboratory, California Institute of Technology, 
  under NASA contract 1407.}

\clearpage
\appendix
\section{Appendix}
\subsection{Calculating rest-frame absolute magnitudes}
The rest-frame absolute magnitude $M_Y$~in a filter $Y$~is calculated
using the general formula
\begin{equation}
M_Y=m_X-5\log(D_L(z)/10 {\rm pc})-K_{XY}(z,T),
\end{equation} 
where $m_X$~is the observed apparent magnitude in filter
$X$,~$D_L(z)$~is the luminosity distance and $K_{XY}(z,T)$ is the
K-correction. The luminosity distance is given by
\begin{equation}
D_L(z) = {c(1+z)\over H_0\ |\Omega_k|^{1/2}} sinn \{ |\Omega_k|^{1/2}
\int_0^x (\Omega_M(1+z^{\prime})^3 + \Omega_k(1+z^{\prime})^2 +
\Omega_{\Lambda})^{-1/2} dz^{\prime}\}
\end{equation} 
where $\Omega_k$ is the curvature term defined as $\Omega_k = 1
-\Omega_M - \Omega_{\Lambda}$ and $sinn$~is defined as $sinh$~for
$\Omega_k > 0$~and $sin$~for $\Omega_k < 0$~(Misner et al. 1973).  If
$\Omega_k = 0$~then $sinn$~and the $|\Omega_k|^{1/2}$-terms are set
equal to one.

We use the formalism in Kim et al. (1996) and Hogg et
al. (2002) to calculate the k-correction at redshift $z$~for a galaxy
template $T$. The generalized equation for calculating the
k-correction is
\begin{equation}
K_{XY}(z,T)=-2.5log_{10}[\frac{1}{(1+z)}\frac{\int d\lambda_o
    \lambda_o f_T(\lambda_o)X(\lambda_o)\int d\lambda_e\lambda_e
    g_{\lambda}(\lambda_e)Y(\lambda_e)}{\int d\lambda_o\lambda_o
    g_{\lambda}(\lambda_o)X(\lambda_o)\int d\lambda_e\lambda_e
    f_T([1+z]\lambda_e)Y(\lambda_e)}],
\end{equation}
where $\lambda_o$~is the wavelength in the observed frame and
$\lambda_e$~is the wavelength in the emitted frame. $X(\lambda)$~and
$Y(\lambda)$ are the filter transmission curves (including corrections
for the detector QE), $f_T$~is the SED of the template $T$ and
$g_{\lambda}$~is the SED of the ``standard source'' used to normalize
the magnitudes. For AB-magnitudes the standard source is a flat
(constant) spectrum in frequency space, $g_{\nu}(\nu)$=constant. In
wavelength space as used here, this corresponds to
$g_{\lambda}(\lambda)=c\lambda^{-2}g_{\nu}(\nu)$.
 
If we know the correct redshift and the correct SED representation of
the observed galaxy, we can calculate the exact rest frame absolute
magnitude with errors only consisting of the observed photometric
errors. However, the true spectral type of the observed galaxy is
generally not known. Instead we represent the the galaxy's SED with
the best-fitting template derived from the photometric redshift
fitting procedure. To minimize the dependence on the SED, we use the
observed filter that best matches the redshifted rest-frame band of
interest, i.e., we want to use an observed filter $X$~that minimizes
\begin{equation}
min\{|\lambda_X - \lambda_Y\times(1+z)|\}
\end{equation}
where $\lambda_X$~is the effective wavelength of the observed filter
and $\lambda_Y$ is the effective wavelength of the rest-frame filter
in which we want to calculate the absolute magnitude. In case of a
perfect match, the k-correction is nearly independent of the assumed
SED, while the dependence on SED increases with the distance between
observed filter and redshifted rest-frame filter.

When calculating
absolute magnitudes in this investigation, we use the two observed
filters $X_a$~and $X_b$~that satisfy equation (A4) and
\begin{equation}
\lambda_{X_a} \le \lambda_Y\times(1+z) < \lambda_{X_b}.
\end{equation}
From the apparent magnitudes in filters $X_a$~and
$X_b$~(i.e. $m_{X_a}$~and $m_{X_b}$), we calculate the corresponding
absolute magnitudes $M_{Y_a}$~and $M_{Y_b}$~ using equation
(A1). Thereafter we interpolate to get $M_{Y}$,
\begin{equation}
M_Y=M_{Y_a}\times(\lambda_{X_b}-\lambda_Y\times(1+z))/(\lambda_{X_b}-\lambda_{X_a})+
M_{Y_b}\times(\lambda_Y\times(1+z)-\lambda_{X_a})/(\lambda_{X_b}-\lambda_{X_a})
\end{equation}
In cases where we do not have two filters available that straddle the
desired rest-frame wavelength, we use the nearest observed magnitude
according to equation (A4), and thereafter set $M_Y$~to either
$M_{Y_a}$~or $M_{Y_b}$, depending on which is available.
\subsection{Errors}
When calculating the LF using the modified $1/V_{\rm max}$~method
(equation 5), the redshift probability distribution for each object
corresponds to a distribution in absolute magnitudes representing the
uncertainty in the latter. However, for many applications it is
desirable to derive an explicit measurement of the error in the
derived rest-frame absolute magnitude. For completeness, we here give
a recipe on how to derive these errors.

Errors in the resulting absolute magnitudes ($\sigma_{\rm tot}$)
mainly come from three sources; 1) photometric errors ($\sigma_m$), 2)
errors due to redshift uncertainty ($\sigma_z$), and 3) errors due to
uncertainty in the best-fitting template ($\sigma_T$). Here we
consider the photometric errors to be known and calculate the
remaining two. The magnitude errors due to uncertainty in the redshift
are largely dominated by the uncertainty in the luminosity
distance. We estimate this error using Monte Carlo simulations where
we recalculate the luminosity distance after adding random errors to
the photometric redshift. The distribution of random errors is assumed
to have a $1\sigma$ dispersion corresponding to the interval
containing 68\% of the redshift probability distribution.

The error due to uncertainty in spectral type is estimated using MC
simulations where we vary the SED when calculating the k-correction in
equation (A1). In the photometric redshift code, we use a number of
SEDs (numbered 1, 2, 3 etc.) where the templates follow an
evolutionary path going from early types to late types. In the
simulations we assign to each object with a nominal best-fitting
template N, a new SED with a random type in the range N-1/2 to
N+1/2. With this `interpolated' type we calculate the absolute
magnitude and estimate variations caused by the change in SEDs. The
Monte Carlo simulations are repeated 10,000 times for each object.

Finally, to get the total error, we add the three parts in quadrature.
\begin{equation}
\sigma_{\rm tot}^2=\sigma_m^2+\sigma_z^2+\sigma_T^2
\end{equation}

\clearpage

\clearpage
\begin{deluxetable}{ccccc}
\tablewidth{0pt}
\tablecaption{Cosmic Variance Estimates}
\tablehead{
\colhead{redshift range} & \colhead{$V_{\rm com}$ (Mpc$^3$)} &
\colhead{$\sigma_{DM}$} & \colhead{$b_{\rm lum}$} & \colhead{$\sigma_L$} 
}
\startdata
\sidehead{\bf WFI ($R$-selected) catalog (area = 1100 arcmin$^2$)}

$0.1<z<0.5$ & $2.07 \times 10^{5}$ & 0.15 & 1.08 & 0.16 \\
$0.5<z<0.75$ & $3.68 \times 10^{5}$ & 0.16 & 1.17 & 0.19 \\
$0.75<z<1.0$ & $5.42 \times 10^{5}$ & 0.15 & 1.24 & 0.19 \\

\sidehead{\bf ISAAC ($K_s$-selected) catalog (area = 130 arcmin$^2$)}

$0.10<z<0.82$ & $8.39 \times 10^{4}$ & 0.15 & 1.12 & 0.16\\
$0.82<z<1.12$ & $8.49 \times 10^{4}$ & 0.21 & 1.27 & 0.27 \\
$1.12<z<1.37$ & $8.65 \times 10^{4}$ & 0.23 & 1.37 & 0.32\\
$1.37<z<1.59$ & $8.44 \times 10^{4}$ & 0.24 & 1.50 & 0.35\\
$1.59<z<1.80$ & $8.54 \times 10^{4}$ & 0.26 & 1.56 & 0.41\\
$1.80<z<2.00$ & $8.42 \times 10^{4}$ & 0.27 & 1.64 & 0.45\\
\enddata
\tablecomments{Table columns are 1) redshift range;  2) comoving volume corresponding to redshift range; 3) variance for dark matter halos; 4) luminosity-weighted effective bias; 5) variance in luminosity density. See Section 3.2 for details.
}
\label{tab:cosvar}
\end{deluxetable}

\clearpage
\begin{deluxetable}{lccccc}
\tabletypesize{\scriptsize}
\tablewidth{0pt}
\tablecaption{Best-fitting Schechter function parameters for rest frame $U$-band luminosity functions}
\tablecolumns{6}
\tablehead{
\colhead{Redshift range} & \colhead{$M^*_U -5\log h_{70}$}& \colhead{$\alpha$}  & \colhead{$\phi^*$}  & \colhead{log($\rho_{\nu}$)}& \colhead{Spectral types}\\ 
\colhead{} & \colhead{} & \colhead{}  & \colhead{10$^{-4}$(Mpc/$h_{70})^{-3}$~mag$^{-1}$}  & \colhead{erg s$^{-1}$~Hz$^{-1}$(h$_{70}$/Mpc$^3$)} & \colhead{}
}
\startdata
\multicolumn{6}{c}{{}}\\
$0.1<z<0.5$ & $-19.98^{+0.08}_{-0.08}$ & $-1.31^{+0.02}_{-0.03}$ & 42.5$^{+3.4}_{-3.8}$  & 26.38$\pm{0.03}$ &All\\
            & $-19.71^{+0.14}_{-0.14}$ & $-0.67^{+0.06}_{-0.06}$ & 14.5$^{+1.6}_{-1.5}$  & &Early-types\\
            & $-19.66^{+0.16}_{-0.16}$ & $-0.64^{+0.09}_{-0.09}$ & 15.1$^{+2.0}_{-1.8}$  & &Early-types$^*$\\
            & $-20.11^{+0.10}_{-0.10}$ & $-1.32^{+0.03}_{-0.03}$ & 23.9$^{+2.9}_{-2.6}$  & &Late-types\\
            & $-18.22^{+0.16}_{-0.20}$ & $-1.03^{+0.14}_{-0.14}$ & 44.7$^{+8.5}_{-9.4}$ & &Starbursts  \\
$0.5<z<0.75$ & $-20.08^{+0.06}_{-0.08}$ & $-1.10^{+0.03}_{-0.04}$ & 64.9$^{+4.5}_{-5.2}$ & 26.51$\pm{0.03}$ &All \\
            & $-19.44^{+0.12}_{-0.10}$ & $-0.17^{+0.11}_{-0.09}$ & 21.8$^{+1.1}_{-1.1}$  & &Early-types\\
            & $-20.24^{+0.08}_{-0.08}$ & $-1.07^{+0.04}_{-0.04}$ & 35.4$^{+3.2}_{-3.2}$  & &Late-types\\
            & $-19.15^{+0.22}_{-0.22}$ & $-1.23^{+0.22}_{-0.20}$ & 32.0$^{+9.4}_{-8.6}$ & &Starbursts  \\
$0.75<z<1.0$ & $-20.32^{+0.10}_{-0.08}$ & $-1.37^{+0.08}_{-0.07}$ & 38.8$^{+5.8}_{-4.7}$ & 26.51$\pm{0.05}$ &All \\
            & $-19.74^{+0.24}_{-0.22}$ & $-0.69^{+0.29}_{-0.26}$ & 9.0$^{+1.5}_{-1.7}$  & &Early-types\\
            & $-20.47^{+0.12}_{-0.10}$ & $-1.26^{+0.10}_{-0.09}$ & 19.2$^{+3.5}_{-2.7}$  & &Late-types\\
            & $-19.71^{+0.18}_{-0.18}$ & $-1.31^{+0.19}_{-0.17}$ & 26.8$^{+7.0}_{-6.2}$ & & Starbursts  \\
\multicolumn{5}{c}{}\\
\multicolumn{5}{c}{{\bf Fit to the same faint-end slope}}\\
\multicolumn{5}{c}{{\it All types, $\alpha=-1.31$}}\\
$0.1<z<0.5$ & $-19.98^{+0.08}_{-0.08}$ & $-1.31^{+0.02}_{-0.03}$ & 42.5$^{+3.4}_{-3.8}$ & 26.38$\pm{0.03}$ &All \\
$0.5<z<0.75$ & $-20.44^{+0.02}_{-0.02}$ & -                       & 40.4$^{+0.4}_{-0.4}$ & 26.54$\pm{0.03}$ &All \\
$0.75<z<1.0$ & $-20.25^{+0.04}_{-0.02}$ & -                       & 43.0$^{+1.7}_{-0.9}$ & 26.49$\pm{0.05}$ &All  \\
\enddata
\tablecomments{(1) $R$-band selected sample.\\
(2) Luminosity densities are derived by integrating LFs characterized by the best-fitting
Schechter function parameters. Errors represent statistical errors. For early-types marked with an asterisk ($^*$),
we have excluded the faint up-turn at $M_U>-16.5$.
}
\label{table1}
\end{deluxetable}

\clearpage
\begin{deluxetable}{lccccc}
\tabletypesize{\scriptsize}
\tablewidth{0pt}
\tablecaption{Best-fitting Schechter function parameters for rest frame $B$-band luminosity functions}
\tablecolumns{6}
\tablehead{
\colhead{Redshift range} & \colhead{$M^*_B -5\log h_{70}$}& \colhead{$\alpha$}  & \colhead{$\phi^*$}  & \colhead{log($\rho_{\nu}$)}& \colhead{Spectral types}\\ 
\colhead{} & \colhead{} & \colhead{}  & \colhead{10$^{-4}$(Mpc/$h_{70})^{-3}$~mag$^{-1}$}  & \colhead{erg s$^{-1}$~Hz$^{-1}$(h$_{70}$/Mpc$^3$)} & \colhead{}
}
\startdata
\multicolumn{6}{c}{{\it Rest-frame B-band}}\\
$0.1<z<0.5$ & $-21.22^{+0.10}_{-0.06}$ & $-1.37^{+0.02}_{-0.01}$ & 28.1$^{+2.8}_{-1.4}$ & 26.73$\pm{0.04}\pm{0.06}$ & All  \\
            & $-21.16^{+0.14}_{-0.12}$ & $-0.74^{+0.05}_{-0.05}$ & 13.1$^{+1.4}_{-1.3}$ &  & Early-types \\
            & $-20.99^{+0.14}_{-0.16}$ & $-0.64^{+0.07}_{-0.07}$ & 15.2$^{+1.5}_{-1.8}$ &  & Early-types$^*$ \\
            & $-21.00^{+0.10}_{-0.08}$ & $-1.35^{+0.02}_{-0.02}$ & 21.2$^{+2.3}_{-1.7}$ &  &Late-types \\
            & $-18.72^{+0.12}_{-0.12}$ & $-1.02^{+0.08}_{-0.08}$ & 42.6$^{+6.0}_{-5.5}$ &  &Starbursts \\
$0.5<z<0.75$ & $-21.46^{+0.06}_{-0.08}$ & $-1.22^{+0.02}_{-0.03}$ & 31.8$^{+2.2}_{-2.9}$ & 26.80$\pm{0.03}\pm{0.07}$ &All \\
            & $-21.01^{+0.14}_{-0.12}$ & $-0.39^{+0.10}_{-0.09}$ & 16.7$^{+1.3}_{-1.3}$ &  &Early-types \\
            & $-21.29^{+0.08}_{-0.08}$ & $-1.14^{+0.04}_{-0.03}$ & 23.4$^{+2.6}_{-2.1}$ &  &Late-types \\
            & $-19.51^{+0.16}_{-0.16}$ & $-1.04^{+0.16}_{-0.14}$ & 27.8$^{+5.3}_{-5.0}$ & &Starbursts \\
$0.75<z<1.0$ & $-21.46^{+0.10}_{-0.12}$ & $-1.24^{+0.08}_{-0.08}$ & 24.5$^{+3.5}_{-3.7}$ & 26.70$\pm{0.05}\pm{0.07}$ &All \\
            & $-21.44^{+0.26}_{-0.28}$ & $-0.72^{+0.28}_{-0.26}$ & 6.2$^{+1.3}_{-1.5}$ &  &Early-types \\
            & $-21.15^{+0.10}_{-0.12}$ & $-0.87^{+0.11}_{-0.12}$ & 22.0$^{+2.4}_{-2.9}$ &  &Late-types \\
            & $-20.32^{+0.20}_{-0.20}$ & $-1.30^{+0.20}_{-0.17}$ & 17.7$^{+5.3}_{-4.4}$ &  &Starbursts \\
\multicolumn{4}{c}{{\bf Fit to the same faint-end slope}}\\
\multicolumn{4}{c}{{\it All types, $\alpha=-1.37$}}\\
$0.1<z<0.5$ & $-21.22^{+0.10}_{-0.06}$ & $-1.37^{+0.02}_{-0.01}$ & 28.1$^{+2.8}_{-1.4}$ & 26.73$\pm{0.04}\pm{0.06}$ &All\\
$0.5<z<0.75$ & $-21.79^{+0.04}_{-0.04}$ & -                       & 20.2$^{+0.4}_{-0.4}$ & 26.81$\pm{0.03}\pm{0.07}$ &All\\
$0.75<z<1.0$ & $-21.62^{+0.04}_{-0.04}$ & -                       & 19.5$^{+0.8}_{-0.6}$ & 26.73$\pm{0.05}\pm{0.07}$ &All\\
\enddata
\tablecomments{(1) $R$-band selected sample.\\
(2) Luminosity densities are derived by integrating LFs characterized by the best-fitting
Schechter function parameters. First errors represent statistical errors, while
second errors represent cosmic variance (see text for details). For early-types marked with an asterisk ($^*$),
we have excluded the faint up-turn at $M_B>-17.5$.
}
\label{table2}
\end{deluxetable}

\clearpage
\begin{deluxetable}{lccccc}
\tabletypesize{\scriptsize}
\tablewidth{0pt}
\tablecaption{Best-fitting Schechter function parameters for rest frame $J$-band luminosity functions}
\tablecolumns{6}
\tablehead{
\colhead{Redshift range} & \colhead{$M^*_J -5\log h_{70}$}& \colhead{$\alpha$}  & \colhead{$\phi^*$}  & \colhead{log($\rho_{\nu}$)}& \colhead{Spectral types}\\ 
\colhead{} & \colhead{} & \colhead{}  & \colhead{10$^{-4}$(Mpc/$h_{70})^{-3}$~mag$^{-1}$}  & \colhead{erg s$^{-1}$~Hz$^{-1}$(h$_{70}$/Mpc$^3$)} & \colhead{}
}
\startdata
\multicolumn{6}{c}{{\it All types}}\\
$0.1<z<0.5$ & $-23.68^{+0.44}_{-0.58}$ & $-1.48^{+0.06}_{-0.05}$ & 7.7$^{+3.7}_{-2.8}$ &27.23$\pm{0.06}$& All  \\ 
            & $-22.97^{+0.64}_{-1.00}$ & $-0.74^{+0.29}_{-0.25}$ & 8.6$^{+4.6}_{-4.2}$ & & Early-types \\
            & $-22.44^{+0.50}_{-0.82}$ & $-0.45^{+0.39}_{-0.33}$ & 12.7$^{+3.8}_{-5.1}$ & & Early-types$^*$ \\
            & $-23.29^{+0.48}_{-0.54}$ & $-1.37^{+0.08}_{-0.06}$ & 8.2$^{+4.2}_{-3.0}$ & & Late-types\\
            & $-20.71^{+0.86}_{-1.00}$ & $-1.65^{+0.31}_{-0.24}$ & 9.4$^{+18.0}_{-7.0}$ & & Starbursts\\
$0.5<z<0.75$ & $-22.88^{+0.12}_{-0.16}$ & $-1.09^{+0.06}_{-0.07}$ & 40.5$^{+6.5}_{-6.9}$ & 27.42$\pm{0.04}$ & All \\
$0.75<z<1.0$ & $-23.09^{+0.24}_{-0.22}$ & $-1.31^{+0.10}_{-0.09}$ & 19.7$^{+6.3}_{-4.9}$ & 27.30$\pm{0.04}$ & All \\
\multicolumn{6}{c}{}\\
\multicolumn{6}{c}{{\bf Fit to the same faint-end slope}}\\
\multicolumn{6}{c}{{\it All types, $\alpha=-1.48$}}\\
$0.1<z<0.5$ & $-23.68^{+0.44}_{-0.58}$ & $-1.48^{+0.06}_{-0.05}$ & 7.7$^{+3.7}_{-2.8}$ & 27.23$\pm{0.06}$ & All \\
$0.5<z<0.75$ & $-23.77^{+0.22}_{-0.22}$ & -                       & 10.4$^{+1.4}_{-1.2}$& 27.40$\pm{0.04}$ & All \\
$0.75<z<1.0$ & $-23.42^{+0.22}_{-0.22}$ & -                       & 12.2$^{+1.1}_{-1.1}$& 27.32$\pm{0.05}$ & All \\
\enddata
\tablecomments{(1) $K_s$-band selected sample.\\
(2) Luminosity densities are derived by integrating LFs characterized by the best-fitting
Schechter function parameters. Errors represent statistical errors.  For early-types marked with an asterisk ($^*$),
we have excluded the faint up-turn at $M_J>-19.0$.
}
\label{table3}
\end{deluxetable}

\clearpage
\begin{deluxetable}{lcccc}
\tabletypesize{\scriptsize}
\tablewidth{0pt}
\tablecaption{Best-fitting Schechter function parameters to rest frame $U$-, $B$- and $R$-band luminosity functions}
\tablecolumns{5}
\tablehead{
\colhead{Redshift range} & \colhead{$M^*_U -5\log h_{70}$}& \colhead{$\alpha$}  & \colhead{$\phi^*$}  & \colhead{log($\rho_{\nu}$)}\\ 
\colhead{} & \colhead{} & \colhead{}  & \colhead{10$^{-4}$(Mpc/$h_{70})^{-3}$~mag$^{-1}$}  & \colhead{erg s$^{-1}$~Hz$^{-1}$(h$_{70}$/Mpc$^{3})$} 
}
\startdata
$0.10<z<0.82$ & $-20.28^{+0.14}_{-0.14}$ & $-1.20^{+0.07}_{-0.05}$ & 54.3$^{+9.2}_{-8.1}$ & 26.55$\pm 0.03$  \\ 
$0.82<z<1.12$ & $-20.26^{+0.10}_{-0.10}$ &  & 55.2$^{+6.1}_{-5.5}$  & 26.55$\pm 0.04$ \\
$1.12<z<1.37$ & $-20.30^{+0.10}_{-0.10}$ &  & 50.5.6$^{+6.6}_{-6.1}$   & 26.53$\pm 0.05$ \\
$1.37<z<1.59$ & $-20.48^{+0.14}_{-0.14}$ &  & 35.8$^{+7.9}_{-6.2}$   & 26.45$\pm 0.06$ \\
$1.59<z<1.80$ & $-20.69^{+0.26}_{-0.26}$ &  & 25.3$^{+14.2}_{-8.9}$ & 26.38$\pm 0.13$ \\
$1.80<z<2.00$ & $-21.16^{+0.36}_{-0.36}$ &  & 18.3$^{+17.9}_{-8.4}$  & 26.43$\pm 0.16$ \\
\hline
\colhead{Redshift range} & \colhead{$M^*_B -5\log h_{70}$}& \colhead{$\alpha$}  & \colhead{$\phi^*$}  & \colhead{log($\rho_{\nu}$)}\\ 
\colhead{} & \colhead{} & \colhead{}  & \colhead{10$^{-4}$(Mpc/$h_{70})^{-3}$~mag$^{-1}$}  & \colhead{erg s$^{-1}$~Hz$^{-1}$(h$_{70}$/Mpc$^{3})$}\\
\hline
$0.10<z<0.82$ & $-21.43^{+0.14}_{-0.13}$ & $-1.28^{+0.05}_{-0.06}$ & 37.9$^{+6.4}_{-6.4}$ & 26.89$\pm 0.04\pm 0.07$  \\ 
$0.82<z<1.12$ & $-21.45^{+0.10}_{-0.10}$ &  & 35.6$^{+2.8}_{-2.8}$  & 26.87$\pm 0.04\pm 0.10$ \\
$1.12<z<1.37$ & $-21.48^{+0.12}_{-0.12}$ &  & 28.6$^{+3.4}_{-3.1}$   & 26.79$\pm 0.05\pm 0.12$ \\
$1.37<z<1.59$ & $-21.60^{+0.16}_{-0.14}$ &  & 21.5$^{+4.1}_{-3.0}$   & 26.71$\pm 0.06\pm 0.13$ \\
$1.59<z<1.80$ & $-21.48^{+0.38}_{-0.40}$ &  & 21.1$^{+22.6}_{-10.1}$ & 26.66$\pm 0.09\pm 0.15$ \\
$1.80<z<2.00$ & $-21.48^{+0.36}_{-0.40}$ &  & 25.5$^{+27.8}_{-13.0}$  & 26.74$\pm 0.14\pm 0.16$ \\
\hline
\colhead{Redshift range} & \colhead{$M^*_R -5\log h_{70}$}& \colhead{$\alpha$}  & \colhead{$\phi^*$}  & \colhead{log($\rho_{\nu}$)}\\ 
\colhead{} & \colhead{} & \colhead{}  & \colhead{10$^{-4}$(Mpc/$h_{70})^{-3}$~mag$^{-1}$}  & \colhead{erg s$^{-1}$~Hz$^{-1}$(h$_{70}$/Mpc$^{3})$}\\
\hline
$0.10<z<0.82$ & $-22.38^{+0.12}_{-0.14}$ & $-1.30^{+0.04}_{-0.05}$ & 28.2$^{+4.2}_{-4.5}$ & 27.15$\pm 0.03$\\ 
$0.82<z<1.12$ & $-22.40^{+0.12}_{-0.10}$ &  & 23.0$^{+1.8}_{-1.4}$  & 27.07$\pm 0.03$ \\
$1.12<z<1.37$ & $-22.38^{+0.14}_{-0.14}$ &  & 18.9$^{+2.1}_{-1.8}$   & 26.98$\pm 0.04$ \\
$1.37<z<1.59$ & $-22.45^{+0.18}_{-0.20}$ &  & 13.9$^{+2.4}_{-2.2}$   & 26.87$\pm 0.05$ \\
$1.59<z<1.80$ & $-22.49^{+0.30}_{-0.34}$ &  & 10.6$^{+4.1}_{-3.0}$ & 26.77$\pm 0.07$ \\
$1.80<z<2.00$ & $-22.23^{+0.40}_{-0.40}$ &  & 13.1$^{+10.3}_{-5.6}$  & 26.36$\pm 0.11$ \\
\enddata
\tablecomments{(1) $K_s$-band selected sample.\\ 
(2) Faint-end slope is fixed to value derived in lowest redshift bin.\\
(3) Luminosity densities are derived by integrating LFs characterized by the best-fitting
Schechter function parameters. Errors in luminosity density represent statistical errors. For the
$B-$band we also represent errors due to cosmic variance (see text for details).
}
\label{table4}
\end{deluxetable}
\end{document}